\documentclass{article}

\usepackage{arxiv}
\usepackage[T1]{fontenc}    
\usepackage{url}            
\usepackage{nicefrac}       
\usepackage{microtype}      
\usepackage{graphicx}
\usepackage[sort&compress,numbers]{natbib}
\usepackage{doi}
\usepackage[utf8]{inputenc}
\usepackage{nomencl}
\usepackage{amsfonts}
\usepackage[bbgreekl]{mathbbol}
\usepackage{etoolbox}
\usepackage{booktabs}
\usepackage{moreverb}
\usepackage{subfig}
\usepackage{tikz,tikzscale}
\usepackage{pgfplots}
\pgfplotsset{compat=1.12}
\usepackage{pgfplotstable} 
\usepackage{pifont}
\usepgfplotslibrary{patchplots}
\usepackage{mathtools}
\usepackage{hyperref}
\usepackage{color}
\usepackage{import}
\usepackage{cleveref}
\usepackage{algorithm}
\usepackage{algpseudocode}
\usepackage{letltxmacro}
\usepackage{rotating}
\usepackage[section]{placeins}
\usepackage[toc,page]{appendix}
\usepackage{amsthm}
\DeclarePairedDelimiterX{\infdivx}[2]{}{}{%
  {#1}\;\big\|\;{#2}%
}

\newcommand{\xtraini}{{\boldsymbol{x}_{i}}}

\newcommand{\Xtrain}{\boldsymbol{X}}
\newcommand{\Xtest}{{{\boldsymbol{X}}^{*}}}
\newcommand{\xtest}{{{\boldsymbol{x}}^{*}}}

\newcommand{\bs}{\boldsymbol}
\newcommand{\bxs}{{\bs{x}_{\text{s}}}}
\newcommand{\bxe}{{\bs{x}_{\text{e}}}}
\newcommand{\bx}{{\bs{x}}}

\newcommand{\xred}{{\boldsymbol{{\hat{x}}}}}
\newcommand{\Xred}{{{\boldsymbol{{\hat{X}}}}^{*}}}
\newcommand{\bg}{\bs{\gamma}}

\newcommand{\gx}{\gamma(\bx)}
\newcommand{\dmfx}{d_{\text{mf}}(\bx)}
\newcommand{\dmf}{d_{\text{mf}}}

\newcommand{\Yobs}{{Y_{\text{obs}}}}

\newcommand{\ylf}{{y_{\text{LF}}}}
\newcommand{\yhf}{{y_{\text{HF}}}}

\newcommand{\ytesthf}{{y_{\text{HF}}^{*}}}

\newcommand{\Ytrainlf}{{Y_{\text{LF}}}}
\newcommand{\Ytrainhf}{{Y_{\text{HF}}}}

\newcommand{\Ytestlf}{{Y_{\text{LF}}^{*}}}

\newcommand{\fmod}{f}
\newcommand{\Ftrain}{F}

\newcommand{\ftest}{{f^{*}}}

\newcommand{\param}{\boldsymbol{\theta}}

\newcommand{\ls}{\ell}
\newcommand{\sv}{{\sigma_{0}^{2}}}
\newcommand{\ops}{{\boldsymbol{z}_{\text{LF}}}}
\newcommand{\opstest}{{\boldsymbol{z}_{\text{LF}}^{*}}}
\newcommand{\Opstest}{{\boldsymbol{Z}_{\text{LF}}^{*}}}
\newcommand{\Opstrain}{{\boldsymbol{Z}_{\text{LF}}}}

\newcommand{\kfun}[3]{\mathrm{k}_{#1}\left(#2,{#3}'\right)}

\newcommand{\nvarraw}{{\sigma_{\text{n}}}^{2}}
\newcommand{\nvar}{{\hat{\sigma}_{\text{n}}}^{2}}

\newcommand{\pdf}[1]{p\left(#1\right)}

\newcommand{\mf}[1]{\mathrm{m}_{\Ds}\left(#1\right)}
\newcommand{\varf}[1]{\mathrm{v}_{\Ds}\left(#1\right)}

\newcommand{\Kmat}{\mathbf{K}}
\newcommand{\Kmattest}{{\mathbf{K}}^{*}}

\newcommand{\ntrain}{\mathrm{n}_{\text{train}}}
\newcommand{\nsample}{\mathrm{N}_{\text{sample}}}
\newcommand{\nfeature}{\mathrm{N}_{\text{feat.}}}

\newcommand{\Ex}[2]{{\mathbb{E}}_{#1}\left[#2\right]}

\newcommand{\Var}[2]{\mathbb{V}_{#1}\left[#2\right]}

\newcommand{\nd}[2]{\mathcal{N}_{#1}\left(#2\right)}
\newcommand{\lnd}[2]{\mathcal{LN}_{#1}\left(#2\right)}
\newcommand{\ud}[2]{\mathcal{U}_{#1}\left(#2\right)}
\newcommand{\GP}[2]{\mathcal{GP}_{#1}\left(#2\right)}

\newcommand{\DKL}[2]{\mathrm{D}_{\text{KL}}\left[#1\big|\big|#2\right]}

\newcommand{\dirac}[2]{\delta_{#1}\left(#2\right)}

\newcommand{\real}[1]{{\mathbb{R}}^{#1}}
\newcommand{\Ds}{\mathcal{D}_{\fmod}}

\newcommand{\Dslfx}{{\mathcal{D}_{\text{LF}}}^{*}}

\newcommand{\Dshfx}{\mathcal{D}_{\text{HF}}}

\newcommand{\ssp}[1]{{\Omega_{{#1}}}}

\newcommand{\opsExt}{\ops^{+}}
\newcommand{\featExt}{{\bs{\gamma}}^{+}}
\newcommand{\FeattestExt}{{\bs{\Gamma}}^{*,+}}
\newcommand{\FeattrainExt}{{\bs{\Gamma}}^{+}}

\newcommand{\OpstrainExt}{\Opstrain^{+}}

\newcommand{\eabs}{\epsilon_{\text{abs}}}

\newcommand{\BMFMC}{\textit{BMFMC }}

\newcommand{\be}{\begin{equation}}
\newcommand{\ee}{\end{equation}}
\newcommand{\bc}{\begin{center}}
\newcommand{\ec}{\end{center}}
\newcommand{\bd}{\begin{description}}
\newcommand{\ed}{\end{description}}
\newcommand{\bi}{\begin{itemize}}
\newcommand{\ei}{\end{itemize}}

\def\RR{ \mathbb{R}}

\def\app#1#2{%
  \mathrel{%
    \setbox0=\hbox{$#1\sim$}%
    \setbox2=\hbox{%
      \rlap{\hbox{$#1\propto$}}%
      \lower1.1\ht0\box0%
    }%
    \raise0.25\ht2\box2%
  }%
}
\def\approxprop{\mathpalette\app\relax}
\DeclareSymbolFontAlphabet{\mathbb}{AMSb}
\DeclareSymbolFontAlphabet{\mathbbl}{bbold}

\definecolor{codegreen}{rgb}{0,0.6,0}
\definecolor{codegray}{rgb}{0.5,0.5,0.5}
\definecolor{codepurple}{rgb}{0.58,0,0.82}
\definecolor{backcolour}{rgb}{0.95,0.95,0.92}

\newlength{\commentindent}
\makeatletter

\LetLtxMacro{\oldalgorithmic}{\algorithmic}
\renewcommand{\algorithmic}[1][0]{%
  \oldalgorithmic[#1]%
}
\makeatother 
\allowdisplaybreaks 

\def\app#1#2{%
  \mathrel{%
    \setbox0=\hbox{$#1\sim$}%
    \setbox2=\hbox{%
      \rlap{\hbox{$#1\propto$}}%
      \lower1.1\ht0\box0%
    }%
    \raise0.25\ht2\box2%
  }%
}
\def\approxprop{\mathpalette\app\relax}

\makeatletter
\newtheoremstyle{indented}
    {3pt}
  {3pt}
  {\addtolength{\@totalleftmargin}{1.5em}
   \addtolength{\linewidth}{-1.5em}
   \parshape 1 1.5em \linewidth}
  {}
  {\bfseries}
  {.}
  {.5em}
  {}
 \makeatother
 \theoremstyle{indented}
\newtheorem*{remark}{Remark}

\title{A Generalized Probabilistic Learning Approach for Multi-Fidelity Uncertainty Quantification in Complex Physical Simulations}

\author{Jonas Nitzler\thanks{corresponding author} \\
	Institute for Computational Mechanics \&\\
	Professorship for Data-driven Materials Modeling\\
	Technical University of Munich\\
	85748 Garching b. München\\
	\texttt{jonas.nitzler@tum.de} \\
	\And
	Jonas Biehler\\
	Institute for Computational Mechanics\\
	Technical University of Munich\\
	85748 Garching b. München\\
	\texttt{jonas.biehler@tum.de} \\
	\AND
	Niklas Fehn\\
	Institute for Computational Mechanics\\
	Technical University of Munich\\
	85748 Garching b. München\\
	\texttt{niklas.fehn@tum.de} \\
	\And
	Phaedon-Stelios Koutsourelakis\\
	Professorship for Data-driven Materials Modeling\\
	Technical University of Munich\\
	85748 Garching b. München\\
	\texttt{p.s.koutsourelakis@tum.de} \\
  Munich Data Science Institute (MDSI)\\
  \url{www.mdsi.tum.de}
	\And
	Wolfgang A. Wall\\
	Institute for Computational Mechanics\\
	Technical University of Munich\\
	85748 Garching b. München\\
	\texttt{wolfgang.a.wall@tum.de} \\
  Munich Data Science Institute (MDSI)\\
  \url{www.mdsi.tum.de}
}

\renewcommand{\shorttitle}{}

\begin{document}
\maketitle

\begin{abstract}
Two of the most significant challenges in uncertainty quantification pertain to the high computational cost for simulating complex physical models and the high dimension of the random inputs. In applications of practical interest, both of these problems are encountered, and standard methods either fail or are not feasible. 
To overcome the current limitations, we present a generalized formulation of a Bayesian multi-fidelity Monte-Carlo~(\emph{BMFMC}) framework that can exploit lower-fidelity model versions in a \emph{small data} regime. The goal of our analysis is an efficient and accurate estimation of the complete probabilistic response for high-fidelity models. \BMFMC circumvents the curse of dimensionality by learning the relationship between the outputs of a reference high-fidelity model and potentially several lower-fidelity models.
While the continuous formulation is mathematically exact and independent of the low-fidelity model's accuracy, we address challenges associated with the \emph{small data regime}~(i.e., only a small number of 50 to 300 high-fidelity model runs can be performed). 
Specifically, we complement the formulation with a set of informative input features at no extra cost. 
Despite the inaccurate and noisy information that some low-fidelity models provide, we demonstrate that accurate and certifiable estimates for the quantities of interest can be obtained for uncertainty quantification problems in high stochastic dimensions, with significantly fewer high-fidelity model runs than state-of-the-art methods for uncertainty quantification.
We illustrate our approach by applying it to challenging numerical examples such as Navier-Stokes flow simulations and fluid-structure interaction problems.
\end{abstract}


\section{Introduction}
\label{sec: intro}
The analysis of complex, real-world systems is usually based on sophisticated, high-fidelity~(HF) computer models. Accuracy comes at the cost of computationally expensive simulations characterized by detailed physical resolution, fine temporal and spatial discretizations, and narrow numerical tolerances. A single evaluation of such models, for example, in large-scale nonlinear and transient biomechanical problems or coupled fluid simulations, can take hours or days, even on modern high-performance clusters. Nevertheless, many questions in industry and science require multiple accurate computer simulations to understand different system configurations or boundary and initial conditions, to perform optimization tasks, or forward and backward uncertainty propagation. Unfortunately, limitations in available resources render the aforementioned types of analysis unfeasible, such that in most practical applications, analysts either avoid such fundamental investigations completely or fall back to less expensive and less accurate lower-fidelity~(LF)\footnote{By \emph{LF model} we mean a lower-fidelity version of the original high-fidelity model which can be generated by, e.g., coarsening of the numerical discretization and solver tolerances and/or by employing a simplified physical description. On the other hand, the term \emph{surrogate} implies a regression model that replaces the physics-based high-fidelity model based on a set of HF solver runs (training data). In case the surrogate predictions are not very accurate, e.g., due to a lack of training data, the surrogate model can also be seen as an LF model in our terminology.} variants of the original model to conduct the analysis.

One strategy to overcome these problems pertains to multi-fidelity schemes, which, by combining information provided by different levels of model sophistication, attempt to decrease the number of high-fidelity model runs of trusted legacy codes while retaining the same accuracy~\cite{ peherstorfer2018survey, Biehler2019}. Especially sampling-based methods for uncertainty propagation, which are often the only choice for nonlinear problems with large variabilities, become unfeasible for costly numerical models. Multi-level Monte-Carlo methods~(MLMC)~\cite{heinrichMultilevelMonteCarlo2001,gilesMultilevelMonteCarlo2008,gilesMultilevelMonteCarlo2015,gilesAdaptiveMultilevelMonte2017} were some of the earliest schemes used to accelerate the calculation of the expectation and variance of a quantity of interest~(QoI) on complex models, given uncertain inputs~$\bx$. The method performs best under a linear dependency between model outputs~(while the method we present in this paper can fully exploit nonlinear model dependencies as well). An estimation of the whole response distribution via MLMC is restricted to special cases~\cite{gilesMultilevelMonteCarlo2015,bierig2016approximation}. A control variate scheme for Monte-Carlo sampling, informed by an LF model, was recently proposed in order to reduce the variance of the statistical estimators~\cite{gorodetsky2020generalized}, hence accelerating the UQ procedure. Other contributions used low-fidelity models to identify important regions in the input space, motivating adaptive sampling strategies and multi-fidelity importance sampling schemes~\cite{li2014adaptive,cui2015data,peherstorfer2016multifidelity}. However, these methods still require costly sampling of the HF model. Similar ideas arose for inverse problems in the form of multi-stage Markov-chain Monte-Carlo methods~\cite{christen2005markov,tierney1999some,fox1997sampling}. 

Alternative methods that exploit LF model information and recently gained considerable attention are so-called
Bayesian multi-fidelity schemes~\cite{Koutsourelakis2009,bilionis2013solution,le2013bayesian,le2014recursive,biehler2015towards,perdikaris2015multi,perdikaris2016model,perdikaris2017nonlinear,quaglino_fast_2018,takhtaganov2018adaptive,Ranftl_2019,grigo2019bayesian,meng2021multi}, especially due to their efficiency in the \emph{small data regime}, i.e., when only a small number of 50 to 300 high-fidelity model runs can be performed. One of the earliest and potentially most influential contributions to this field was already published in 2000 by Kennedy and O`Hagan~\cite{kennedy2000predicting}. Similar to most state-of-the-art, multi-fidelity methods, the approach aims to construct an approximation for an HF output~$\yhf(\bx)$ in the form of an LF or surrogate model~$\hat{f}(\bx)\approx\yhf(\bx)$  based on few HF simulations. These multi-fidelity approaches exploit that the discrepancy between the HF and LF model response has a simpler mathematical structure than the HF model response~$\yhf(\bx)$ itself. Hence, the discrepancy over~$\bx$ can be efficiently learned to yield good HF response predictions using few data~\cite{kennedy2000predicting,le2013bayesian,le2014recursive,perdikaris2015multi,perdikaris2017nonlinear}. Unfortunately, such surrogate-based approaches face severe problems in applications with high stochastic dimension~($\dim(\bx)>10$), especially in the case of a \emph{small data} scenario, which refers to the small number of HF simulations available due to the associated high costs. 

Even though not further addressed in this paper, we also want to note the recent developments in the field of~(physics-informed) neural networks~\cite{raissi2017inferring,yang2019adversarial,meng2020composite,mahmoudabadbozchelou2021data,yang2021b}. The latter solve an underlying PDE by physics-constrained deep learning methods, potentially simultaneously over a respective parameterization. In contrast to classical surrogate approaches, which substitute the high fidelity solver with a computationally cheap to evaluate regression model constructed upon a set of HF solver runs, physics-informed neural networks do not require HF training data provided by computationally demanding solver runs. Instead, they optimize a loss function that enforces the physical PDE. The physical guidance allows mitigating the curse of dimensionality that arises from high dimensional parameter spaces. Given a physics-informed neural network, UQ can then be performed by sampling the latter for the uncertain parameters. While these ideas offer an interesting pathway, they are so far not yet able to replace trusted legacy codes in combination with large-scale complex and coupled mechanical models, in which we are interested in this work.

In this paper, we want to present an advancement on \emph{Bayesian multi-fidelity Monte-Carlo (BMFMC)}, initially developed by one of the authors in~\cite{Koutsourelakis2009}. \BMFMC is a non-intrusive, data-driven approach that can be used with any numerical solver. It is not a~(multi-fidelity) surrogate-based approach that approximates a mapping from the input space to the output space. It can circumvent the curse of dimensionality imposed by high-dimensional stochastic inputs~$\bx$. Instead, it achieves its superior performance in high-stochastic dimensions by learning a statistical and potentially nonlinear dependency between the low-dimensional output of a high-fidelity and a low-fidelity model. Point estimates such as event probabilities or expectations can then be calculated along with credible intervals due to the Bayesian nature of our approach.
We already demonstrated the performance of \emph{BMFMC} for large-scale biomechanical problems in high stochastic dimensions~\cite{biehler2015towards}. 

We expand these ideas by presenting a generalized version of \BMFMC in the present contribution:
Specifically, we address the challenges imposed by the modeling error due to a specific choice of a discriminative model and challenges imposed by the epistemic uncertainty due to a very limited amount of high-fidelity model runs. We propose a strategy to minimize these error sources by employing a low-dimensional set of informative input features combined with the LF model(s) at no additional computational cost. This leads to significant accuracy gains in the method, which allows us to exploit a broader range of automatically generated low-fidelity versions of the original problem. Throughout the paper, we present a general theoretic viewpoint on Bayesian uncertainty propagation while emphasizing the practical applicability of the proposed techniques to a broad field of engineering problems. 

The paper is structured as follows: 
In \Cref{sec: general_remarks}, we present the theoretical foundation for Bayesian multi-fidelity uncertainty quantification. After the general presentation in Section \ref{sec: multi_fidelity_overview}, we focus in Section \ref{sec: probabilistic_model} on the approximation in the \emph{small data regime} with the use of a discriminative model. In Section \ref{sec: numerical_approx} we derive the posterior statistics for the particular choice of Gaussian Process Regression as a discriminative model.
Section \ref{sec: informative_features} is then devoted to the meaning and computation of informative input features~$\gamma_i(\bx)$, which represent a crucial contribution to this paper and increase the accuracy of the method. The theoretical part of this paper concludes with an algorithmic summary in Section \ref{sec: algorithmic_summary} followed by a short analysis of computational complexity and speed-up in Section \ref{sec: aspects_implementation}.
The performance of the proposed methodological framework is demonstrated in \Cref{sec: numerical_demonstration} on two relevant problems, namely a fluid mechanics and a fluid-structure interaction problem. Apart from the algorithmic aspects, we also focus on modeling physically compatible random boundary conditions, random fields, and their numerical realization. 
We conclude with a discussion of the numerical results and computational performance and provide an outlook on possible future developments in Section \ref{sec:conclusion}.

\section{Bayesian Multi-Fidelity Uncertainty Quantification}
\label{sec: general_remarks} 

Uncertainty quantification~(UQ) aims to propagate the uncertainty of a random input vector~$\bx\in\Omega_{\bx}\in\RR^d$, with a given density~$\pdf{\bx}$ through a physics-based, high-fidelity, numerical model to accurately and efficiently quantify the uncertainty of one or more outputs or quantities of interest~$\boldsymbol{y}$, for example in the form of their density~$\pdf{\yhf}$. The random vector~$\bx$ can represent uncertainties in model parameters, loads, excitations, or boundary / initial conditions. For applications of practical interest, its dimension~$\dim(\bx)$ is very high~(in the hundreds or thousands). On the other hand, in backward uncertainty propagation, given a similar mathematical model and, in general, noisy observations~$\Yobs$ of the system's output~$\bs{y}$, the goal is to estimate a vector of the model inputs~$\bx$.

In the following, we denote by~$\yhf(\bx)$ the \emph{deterministic} input-output map implied by a high-fidelity model, which in most cases of practical interest is not available in closed form and expensive to evaluate~(e.g., for each value of~$\bx$ the numerical solution of time-dependent, nonlinear PDEs needs to be carried out).
We assume that the high-fidelity model is the reference model, i.e., its predictions~$\yhf$ coincide with the QoI. For clarity of the presentation we consider the scalar case, i.e.,~$\yhf: \Omega_{\bx} \to \RR$. Furthermore, to simplify the notation, we make no distinction between random variables and the values these can take. In this notation,~$\bx$ or~$\yhf$ denote the respective random variables and possible realizations, whereas~$\yhf(\bx)$ refers to a deterministic function. Plain letters express scalar quantities~(e.g.,~$\yhf$ for a scalar, high-fidelity model output), in contrast to boldface letters~(such as the input vector~$\bx$), which denote vector-valued quantities. We denote with capital letters a data set that can either consist of scalar quantities or vector-valued quantities. Data sets of scalar quantities are written in plain capital letters, such as the vector of row-wise scalar experimental observations~$\Yobs$. In contrast, vector-valued quantities, such as the matrix of row-wise vector-valued model inputs~$\Xtrain$, are written with boldface capital letters. Furthermore, a distinction is made between training data of a probabilistic model~(indicated by capital letter but without further superscripts, e.g.,~$\Ytrainhf$) and test data of a probabilistic model, that has an asterisk superscript~(e.g.,~$\xtest$ for one arbitrary test input or the large data set of all test inputs~$\Xtest$).

We seek the whole response density~$\pdf{\yhf}$ which can then be used to calculate any statistic of interest. The resulting output density for the QoI can be expressed as the integral over the conditional distribution~$\pdf{\yhf|\bx}$ weighted by the density of the input~$\pdf{\bx}$. In the case of a deterministic function~$\yhf(\bx)$, the conditional distribution~$\pdf{\yhf|\bx}$ can be expressed in form of a Dirac distribution~$\pdf{\yhf|\bx}=\dirac{\yhf}{\yhf-\yhf(\bx)}$:
\begin{equation}
    \label{eqn: forward_UQ}
    \pdf{\yhf} = \int_{\Omega_{\bx}}\pdf{\yhf|\bx}\pdf{\bx} d \bx=\int_{\Omega_{\bx}}\dirac{\yhf}{\yhf-\yhf(\bx)}\pdf{\bx} d \bx
\end{equation}
\Cref{eqn: forward_UQ} is usually approximated by Monte-Carlo methods which, depending on the desired level of accuracy and especially at the tails~(i.e., rare events) of~$\pdf{\yhf}$, typically require many evaluations of the HF model~$\yhf(\bx)$. The overall computational cost can render such an approach impracticable or unfeasible. Alternative strategies have attempted to approximate the map~$\yhf(\bx)$ or the conditional density~$\pdf{\yhf|\bx}$ using a variety of surrogates or emulators which are generally trained on~$\ntrain$ simulation data pairs, i.e.,~$\Dshfx=\{\xtraini,\yhf(\xtraini)\}_{i=1}^{\ntrain}$. Given the high dimension~$d$ of~$\bx$, this task gives rise to several accuracy and efficiency challenges. Even when the most expressive, modern machine learning tools are deployed~(e.g., Deep Neural Nets), the number~$\ntrain$ of  high-fidelity training evaluations needed to achieve an acceptable level of accuracy can render such methods impracticable or unfeasible as well.

\subsection{General Aspects of the Bayesian Multi-Fidelity Formulation}
\label{sec: multi_fidelity_overview}

The previous expression~\eqref{eqn: forward_UQ} involved a computationally expensive high-fidelity computer model implied by~$\yhf(\bx)$. In the following, we demonstrate how less expensive lower-fidelity models combined with low-dimensional features of~$\bx$ can be employed to obtain accurate and certifiable estimates of the quantities above. In the simplest version we presuppose the availability of a lower-fidelity model, which provides a potentially very poor approximation of the QoI. We denote its output with~$\ylf$ and the associated input-output~(deterministic) map by~$\ylf(\bx)$.

\begin{figure}[!htb]
 \centering
  \includegraphics[scale=0.3]{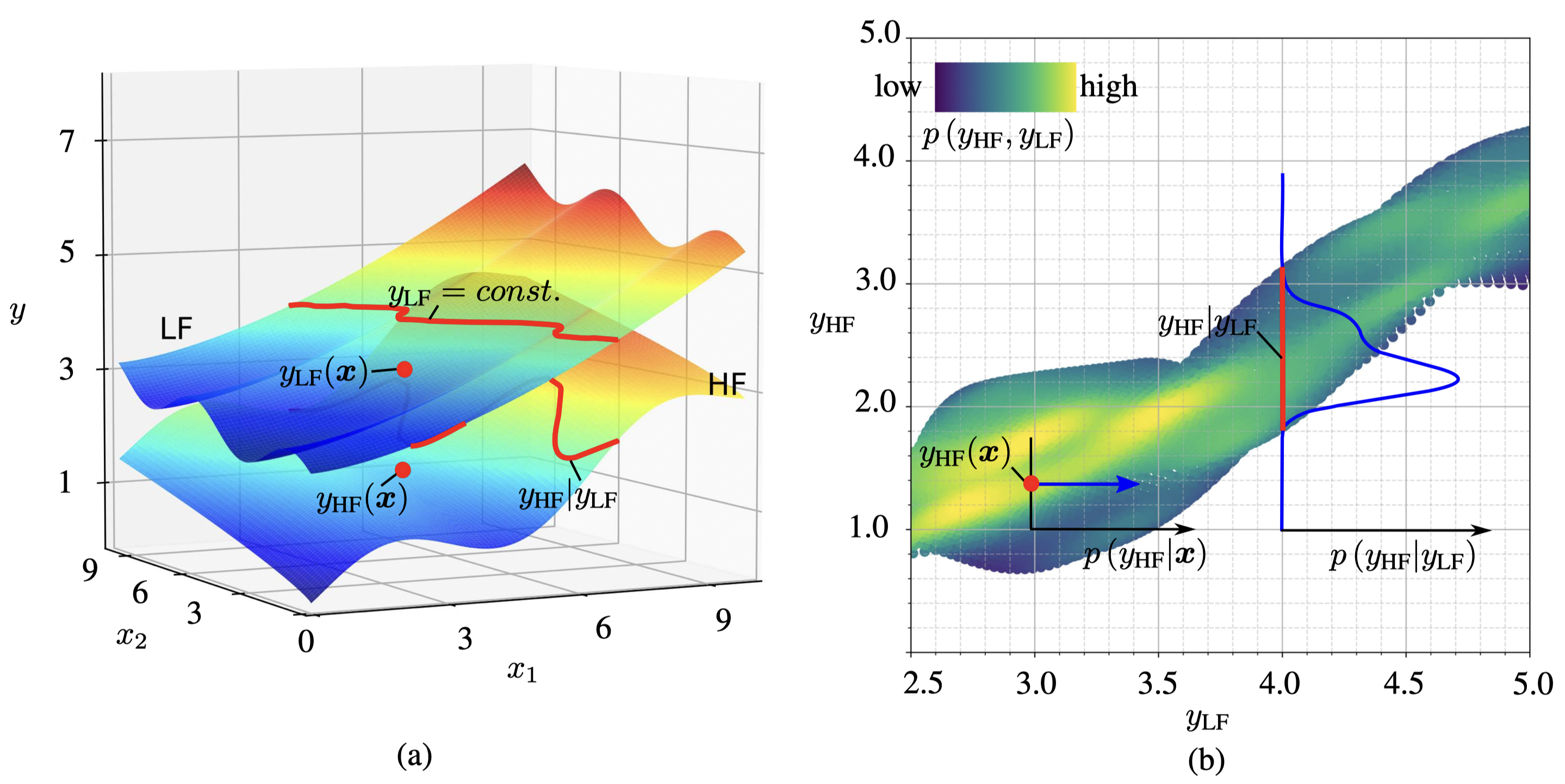}
  \caption{Visualization of HF and LF model dependencies. Left: Example of LF and HF model outputs with two input variables~$x_1$ and~$x_2$. Right: Dependence between LF and HF output. The joint density~$\pdf{\ylf,\yhf}$ is color-coded. Conditional densities~$\pdf{\yhf|\bx}$ and~$\pdf{\yhf|\ylf}$ are shown as slices of~$\pdf{\ylf,\yhf}$ in blue.} 
  \label{fig: conditionals}
 \end{figure}

Contrasting multi-level Monte-Carlo techniques, which also make use of lower-fidelity models in combination with frequentist estimators, we advocate a Bayesian perspective~\cite{Koutsourelakis2009}, which we refer to as \emph{Bayesian Multi-Fidelity Monte-Carlo}~(BMFMC) method~\cite{biehler2015towards,quaglino_fast_2018}. The basis of the framework is re-expressing the sought density as:
\begin{align}
\label{eqn: mf_uq}
\pdf{\yhf}&=\int\limits_{\ssp{\bx}} \underbrace{\pdf{\yhf|\bx}}_{\substack{\text{Dirac:}\\ \text{comp. expensive}}} \cdot\pdf{\bx}~d\bx&&\rightarrow\text{standard forward UQ}\nonumber\\[1.0ex]
&=\int_{\ssp{\ylf}}\int_{\ssp{\bx}} \pdf{\yhf,\ylf,\bx}d\bx d\ylf &&\rightarrow\text{expand by LF model}\\[1.0ex]
&=\int_{\ssp{\ylf}}\int_{\ssp{\bx}}  \pdf{\yhf,\bx|\ylf}\cdot\pdf{\ylf}d\bx d\ylf&&\rightarrow\text{condition on LF model}\nonumber\\[1.0ex]
&=\int_{\ssp{\ylf}} \underbrace{\pdf{\yhf|\ylf}}_{\substack{\text{Approximate with}\\ \text{little HF data}}}\cdot\underbrace{\pdf{\ylf}}_{\substack{\text{Sampling}\\ \text{on LF}}} d\ylf&&\rightarrow\text{integrate over inputs~$\bx$}\nonumber
\end{align}
We note that none of the expressions above contain \emph{any errors or approximations}. Furthermore, the crucial conditional density~$ \pdf{\yhf |  \ylf} $  that must be learned or estimated is \emph{independent of the dimension of the input vector~$\bx$}. 
The premise of \emph{BMFMC} is that all the densities above can be estimated at a cost~(as measured by the number of high-fidelity solves) that is much less than the alternatives. In the case of~$\pdf{\ylf}$ this can be achieved as long as the lower-fidelity model is much cheaper than the high-fidelity reference. To assess the feasibility of this task for~$ \pdf{\yhf |  \ylf} $ and to better understand the role of this conditional density, we consider the following limiting cases:
\begin{description}
\item[extreme 1)] The LF model is independent of~$\yhf$, i.e.,~${\pdf{\yhf |  \ylf}=\pdf{\yhf}}$. While~\eqref{eqn: mf_uq} remains valid, any attempt to estimate  
~$\pdf{\yhf |  \ylf}$ will be comparable to a Monte-Carlo estimator applied directly on~$\yhf$. Hence, it is unlikely that any significant efficiency gains could be achieved.
 \item[extreme 2)] The LF and HF model are fully dependent, i.e., there is a function~$\fmod$, such that~$\yhf=\fmod(\ylf)$ and~$\pdf{\yhf |  \ylf}=\dirac{\yhf}{\yhf-\fmod(\ylf)}$. Any efficiency gains would depend on the cost of learning~$f$, e.g., in terms of HF model runs.
\end{description}
In realistic settings, one would expect the actual~$\pdf{\yhf | \ylf}$ to be between these two extremes as shown in \Cref{fig: conditionals}. While Equation \eqref{eqn: mf_uq} still assumes that the multi-fidelity conditional $\pdf{\yhf|\ylf}$ is known exactly, the aforementioned \emph{realistic settings} require an approximation of the latter based on limited data $\Ds$, as demonstrated in more detail in Section~\ref{sec: probabilistic_model}. Due to our formulation's fully Bayesian nature, the multi-fidelity approximation of the HF output density $\pdf{\yhf}$ is given as a random process itself. We can derive a point estimate $\pdf{\yhf|\Ds}$ (which is a full density function), as an approximation for the HF output density, along with credible bounds for our estimate (see Section \ref{sec: probabilistic_model}). The term \emph{error} describes in this context a deviation of our point estimate $\pdf{\yhf|\Ds}$ from the true (but usually unknown) HF density $\pdf{\yhf}$. Errors will only be introduced through the numerical approximation of the densities in \Cref{eqn: mf_uq}. We distinguish between the following two error sources:
\begin{description}
\item[error 1)] The first error source pertains to the probabilistic model~(e.g., Gaussian Processes or other probabilistic regression tools) selected to approximate the conditional density~$\pdf{\yhf | \ylf}$. If the family of approximating densities considered does not include the true one, a modeling error will be introduced. This error might affect all output statistics, including the credible bounds of the prediction. For example, if a Gaussian Process is employed, only Gaussian conditionals~$\pdf{\yhf|\ylf}$ can be captured. Additional limitations arise from the covariance kernel(s) adopted.
\item[error 2)] The second error source pertains to the amount of available training data~(i.e., pairs of lower- and high-fidelity runs). Even if the true~$\pdf{\yhf|\ylf}$ can be described precisely by the probabilistic model, it is unlikely to be recovered exactly with finite data. On the one hand, the training of the probabilistic regression approach is more likely to deteriorate in a small data regime. On the other hand, the increasing epistemic uncertainty leads to wider credible bounds in the multi-fidelity prediction for the HF output density, rendering a point estimate inconclusive. Nevertheless, we also want to emphasize that the ability to actually provide such credible bounds is a strength of the proposed method.

\end{description}
Please note that we assume that the error in estimating~$\pdf{\ylf}$ can be made arbitrarily small due to the low cost of LF simulations. \emph{BMFMC} brings two significant advantages for UQ with computationally demanding computer models: 
Firstly, by exploiting information encoded in computationally cheaper, lower-fidelity~(LF) versions of the original computer model, we can drastically reduce the number of costly HF model evaluations and enable forward and backward uncertainty propagation even for very expensive models.
Secondly, by learning a nonlinear statistical dependency between the~$\ylf$ and~$\yhf$, \emph{BMFMC} circumvents the curse of dimensionality, which arises as a result of high-dimensional model inputs~$\bx$. 

\Cref{fig: conditionals} provides a visualization of the main terms in Equation~\eqref{eqn: mf_uq}. In particular,  \Cref{fig: conditionals}(a) shows examples of  response surfaces for a two-dimensional input~$\bx$. The upper response surface represents an LF model, and the lower one the corresponding HF model response. A red dot marks a function value for the same~$\bx$ on both models. 
The corresponding Dirac density~$\pdf{\yhf|\bx}$ is shown by a blue arrow, centered on the red dot in \Cref{fig: conditionals}(b).  An indicative conditional density of~$\yhf$ given~$\ylf$ is also shown in \Cref{fig: conditionals}(b).
The vertical red line shows the support for the corresponding conditional density~$\pdf{\yhf|\ylf}$ which encodes the knowledge about possible outcomes of~$\yhf$ when only~$\ylf$ is known~(without information of a specific~$\bx$ that yielded~$\ylf$). 

In this contribution we generalize expression~\eqref{eqn: mf_uq} by considering, in addition to LF models, informative features~$\bs{\gamma}(\bx)=\{\gamma_j(\bx)\}_{j=1}^{\nfeature}$ of the input~$\bx$, with~$\nfeature$ being the number of informative input features used in the multi-fidelity approach. We will further elaborate on this in \Cref{sec: informative_features}. We denote the vector of informative input features~$\bs{\gamma}(\bx)$ and the LF response~$\ylf(\bx)$ as~$\ops(\bx)=[\ylf(\bx),\bs{\gamma}(\bx)]^T$. With~$\ops$, we denote jointly the corresponding random vector as well as values that this can take. The basic elements of \emph{BMFMC} remain unaltered if one employs multiple low-fidelity features, summarized in~$\ops$, so that Equation~\eqref{eqn: mf_uq} becomes:
\be
\pdf{\yhf}=\int\limits_{\ssp{\ops}} \pdf{\yhf | \ops}\cdot \pdf{\ops} d\ops
\label{eqn: uq_extended}
\ee
We demonstrate in the subsequent sections how the modeling error~(i.e., \textbf{error 1} above) can be reduced and how superior estimates can be obtained by an appropriate selection of the input features.

\begin{figure}[!htb]
 \centering
  \includegraphics[scale=0.7]{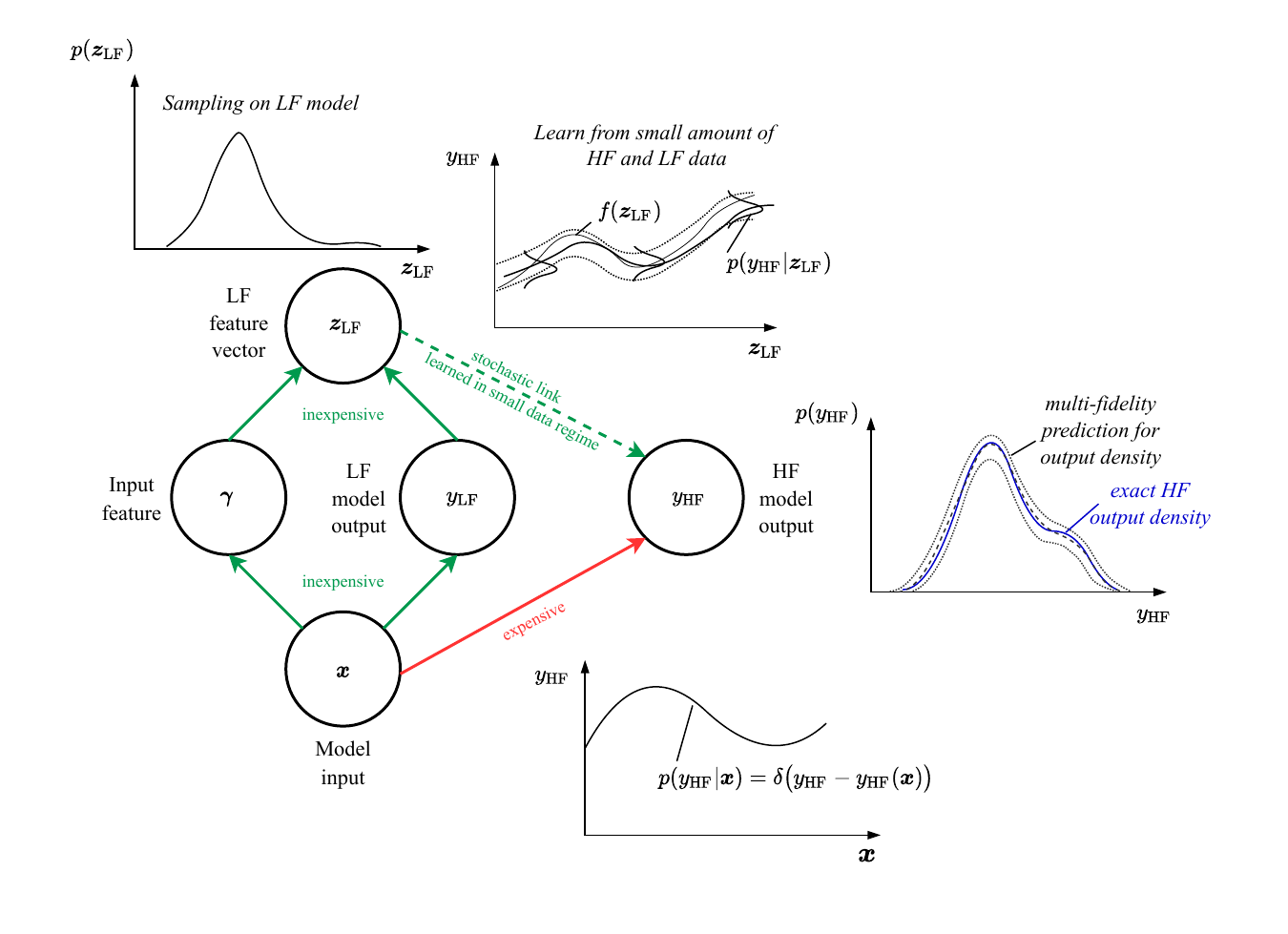}
  \caption{Connection between the involved quantities in the generalized formulation of \BMFMC. Instead of solving the expensive HF model~(red arrow between~$\bx$ and~$\yhf$) repeatably, we follow the computationally cheaper, green path over the LF model output~$\ylf$ and the input feature vector~$\bg$. Together, the two latter variables form the LF feature vector~$\ops$. Its density can be accurately and inexpensively approximated with direct Monte-Carlo, i.e., by drawing samples~$\bx\sim \pdf{\bx}$ and evaluating the LF model and the informative input features. Together with the multi-fidelity conditional distribution~$\pdf{\yhf|\ops}$, which we learn from a small amount of training data~$\Ds$, we can then make predictions for the HF output density~$\pdf{\yhf}$. The link between~$\ops$ and~$\yhf$ is stochastic and hence depicted by a dashed line. \BMFMC results in a stochastic description of~$\pdf{\yhf}$ due to the epistemic uncertainty introduced by the approximation of the multi-fidelity conditional~$\pdf{\yhf|\ops}$.} 
  \label{fig: variables_visu}
 \end{figure}

An overview of the generalized \BMFMC formulation and the connections between the involved quantities is given in Figure \ref{fig: variables_visu}.

 \FloatBarrier
\subsection{Learning of~$\pdf{\yhf|\ops}$: Multi-Fidelity UQ in the Small Data Case}
\label{sec: probabilistic_model}

Only a few HF model evaluations can be afforded in real applications due to their computational expense. Hence, this section focuses on strategies to efficiently learn the multi-fidelity conditional~$\pdf{\yhf|\ops}$ which is the key element of the multi-fidelity approach. Apart from the obvious accuracy requirements, it is essential that the necessary number of training data~$\Ds=\{\ops(\bx^{(i)}) , \yhf(\bx^{(i)}) \}_{i=1}^{\ntrain}$, is minimized.

To approximate the multi-fidelity density~$\pdf{\yhf|\ops}$ in the small data case, we propose the following two steps. Firstly, we make use of a discriminative model that attempts to express the conditional distribution~$\pdf{\yhf|\ops}$ by a functional relationship between~$\yhf$ and~$\ops$. Particularly, a discriminative model learns a~(posterior) distribution over functions~$f(\ops)$ from the LF and HF data set~$\Ds$. For a specific test point~$\opstest$ we can then write the associated distribution of the function value~$\ftest$ as~$\ftest\sim\pdf{\ftest|\opstest,\Ds}$. In Section \ref{sec: numerical_approx} we will show how this distribution can be learned from data~$\Ds$ and incorporated into our formulations for the specific choice of a Gaussian Process regression model. 
Secondly, we postulate a noise model for the HF model output with respect to~$\ftest$ which is expressed by~$\ytesthf\sim\pdf{\ytesthf|\opstest,\ftest}$. The expression might simplify to~$\ytesthf\sim\pdf{\ytesthf|\ftest}$ if the noise~(model) is assumed to be independent of the LF vector~$\opstest$. We can summarize the model for the multi-fidelity conditional~$\pdf{\yhf|\ops}$, which is now additionally dependent on~$\ftest$, by:
\begin{align}
\pdf{\yhf|\ops}\rightarrow\pdf{\ytesthf|\ftest,\opstest}\ \text{with }\ftest\sim\pdf{\ftest|\opstest,\Ds}
\label{eqn: extended_mf}
\end{align}

The model for the multi-fidelity conditional from Equation~\eqref{eqn: extended_mf} can now be plugged into Equation~\eqref{eqn: mf_uq} to yield a multi-fidelity UQ formulation which is  dependent on~$\ftest$:
\begin{equation}
\label{eqn: hl_BMFMC1}
\pdf{\ytesthf|\ftest}=\int\limits_{\ssp{\ops}} \underbrace{\pdf{\ytesthf|\ftest,\opstest}}_{\mathclap{\substack{\text{Likelihood of}\\ \text{HF observations}}}} \cdot \underbrace{\pdf{\ops}}_{\mathrlap{\substack{\text{Marginal density:}\\ \text{direct MC on LF model}}}} d \ops
\end{equation}
As we are not interested in specific values of the regression function~$\ftest$, we will eliminate the dependency on the latter by calculating statistics of the HF distribution over the random variable~$\ftest$. 

Specifically, we are interested in the expectation and variance of Equation~\eqref{eqn: hl_BMFMC1} concerning the random variable~$\ftest$. Under some slight abuse of notation, we write the random variable for which the statistics are computed as a subscript to the respective operator symbol. The expectation~${\Ex{\ftest}{\pdf{\ytesthf|\ftest}}}$ serves as an approximation for the HF distribution~$\Ex{\ftest}{\pdf{\ytesthf|\ftest}}\approx\pdf{\yhf}$. Our confidence in this prediction can be expressed in the form of the variance of Equation~\eqref{eqn: hl_BMFMC1} with respect to~$\ftest$. The variance estimate motivates credible intervals for the density function~$\Ex{\ftest}{\pdf{\ytesthf|\ftest}}$ itself.
The expectation and the variance of the HF output density~$\pdf{\ytesthf|\ftest}$ with respect to~$\ftest$ are given below. Note, that Equations \eqref{eqn: hl_BMFMC2} and \eqref{eqn: hl_BMFMC3} arise from Equation \eqref{eqn: uq_extended} after employing a specific discriminative model for the multi-fidelity conditional (see Equation \eqref{eqn: extended_mf}) in the small data regime.
\begin{subequations}
\begin{align}
\label{eqn: hl_BMFMC2}
\begin{split}
\underbrace{\Ex{\ftest}{\pdf{\ytesthf|\ftest}}}_{\approx \pdf{\yhf}}&=\int\limits_{\ssp{\opstest}}\underbrace{\int\limits_{\ssp{\ftest}} \pdf{\ytesthf|\ftest,\opstest}\pdf{\ftest|\opstest,\Ds}d\ftest}_{\Ex{\ftest}{\pdf{\ytesthf|\ftest,\opstest}}\approx\pdf{\yhf|\ops}} \pdf{\opstest} d\opstest
\end{split}\\
\label{eqn: hl_BMFMC3}
\begin{split}
\Var{\ftest}{\pdf{\ytesthf|\ftest}}&=\Ex{\ftest}{{\left(\pdf{\ytesthf|\ftest}\right)} ^{2}}-{\left(\Ex{\ftest}{\pdf{\ytesthf|\ftest}}\right)} ^{2}\\
&=\int\limits_{\ssp{\opstest}}\int\limits_{{\ssp{\opstest}}'}\Ex{\ftest}{\pdf{\ytesthf|\ftest,\opstest}\pdf{\ytesthf|\ftest,{\opstest}'}}\\
&\quad\cdot\pdf{{\opstest}'}\pdf{\opstest}d{\opstest}'d\opstest-{\left(\Ex{\ftest}{\pdf{\ytesthf|\ftest}}\right)} ^{2}
\end{split}
\end{align}
\end{subequations}

\subsection{Numerical Approximation of Posterior Statistics using Gaussian Processes}
\label{sec: numerical_approx}

In this work, we advocate  Gaussian Processes~(GPs)~\cite{CarlEdwardRasmussen2005} as a probabilistic regression model in Equation~\eqref{eqn: extended_mf} but want to emphasize that any other choice of discriminative model is possible as well. GPs are a popular non-parametric Bayesian tool that is well-suited to small data settings. For the inference and learning tasks, we employed \emph{GPy}~\cite{gpy2014}. The interested reader is referred to Appendix \ref{appendix: gp_approx} for details concerning the specifics of the GP model and set-up. Given  a  test input~$\opstest$, the predictive posterior  of the value of the GP at this point, i.e.,~$\ftest(\opstest)$, is given by a normal distribution:
\begin{align}
\label{eqn: test_input_GP}
\begin{split}
\pdf{\ftest|\opstest,\Ds}&=\GP{\ftest}{\mf{\opstest},\mathrm{k}_{\Ds}\left( \opstest, \opstest \right)}\\
&=\nd{\ftest}{\mf{\opstest},\varf{\opstest}}
\end{split}
\end{align}
Furthermore, the Gaussian likelihood~(see Equation~\eqref{eqn: likelihood} in Appendix \ref{appendix: gp_approx}) implies that the predictive distribution for the corresponding value of the HF model's output, denoted by~$\ytesthf$, will be:
\be
\pdf{\ytesthf | \ftest,\opstest}=\pdf{\ytesthf | \ftest}=\nd{\ytesthf}{\ftest, \nvar}
\label{eq:pr}
\ee 
Here, $\nvar$ is an optimized hyper-parameter of the GP, representing the average variance of the predictive HF output $\ytesthf$ with respect to a realization of the GP $\ftest$. As pointed out in more detail in Appendix \ref{appendix: gp_approx}, we determine a point estimate $\hat{\param}$ for all hyper-parameters $\param=\{\ls\ , \sv\ , \sigma_n^2 \}$ of the employed GP model by maximizing its marginal likelihood with respect to $\param$~\cite{CarlEdwardRasmussen2005}. The remaining hyper-parameters $\ls$ and $\sv$ represent the length-scale, respectively, the signal variance of the squared exponential kernel function used in this paper (see Appendix \ref{appendix: gp_approx} for further information).

Based on the previous results and given the posterior uncertainty of the GP, we can compute the  expected value of the density~$\Ex{\ftest}{\pdf{\ytesthf|\ftest}}$ in Equation~\eqref{eqn: hl_BMFMC2}, by averaging over the posterior of the GP:
\begin{equation}
\label{eqn: cond_expectation}
\Ex{\ftest}{\pdf{\ytesthf|\ftest}}\approx \frac{1}{\nsample}\sum\limits_{j=1}^N \nd{\ytesthf}{\mf{\opstest^{(j)}},\varf{\opstest^{(j)}}+\nvar}
\end{equation}
We note, that the Monte-Carlo approximation for the involved integrals depends on inexpensive LF samples~$\Opstest=[\opstest_i]$\ with~$i\in\mathbb{N}:[1,\nsample]$. For the posterior mean~$\mf{\opstest}$ and variance~$\varf{\opstest}$ of the GP, no additional HF runs are needed. The detailed derivation of~\eqref{eqn: cond_expectation} is given in Appendix \ref{appendix: gp_approx} in Equation~\eqref{eqn: cond_expectation_refined}.

Similarly, the~(posterior) variance of the approximation to the sought density~$\pdf{\yhf}$ can be computed from~\eqref{eqn: hl_BMFMC3} by substituting the GP approximations~(see Appendix \ref{appendix: gp_approx} and in particular Equation~\eqref{eqn: cond_variance_refined} and \Cref{tab: density_models} for more details) and using again Monte-Carlo integration and already computed LF samples~$\opstest_i$: 
\begin{align}
    \label{eqn: cond_variance}
    \begin{split}
\Var{\ftest}{\pdf{\ytesthf|\ftest}}&\approx\frac{1}{\nsample^2}\sum\limits_{i,j=1}^{\nsample}
\nd{\mathbf{\ytesthf}}{\begin{bmatrix}
\mf{\opstest^{(i)}}\\
\mf{\opstest^{(j)}}
\end{bmatrix},\begin{bmatrix}
\varf{\opstest^{(i)}}+\nvar & \kfun{\Ds}{\opstest^{(i)}}{\opstest^{(j)}} \\
\kfun{\Ds}{\opstest^{(j)}}{\opstest^{(i)}} & \varf{\opstest^{(j)}}+\nvar
\end{bmatrix}}\\
&\quad-\left(\Ex{\ftest}{\pdf{\ytesthf|\ftest}}\right)^2\\
\end{split}
\end{align}
In the following section we will now discuss the details and the composition of~$\ops$ and additionally provide pseudo-algorithms to summarize all essential steps of the Bayesian multi-fidelity approach before we provide some considerations on computational complexity and numerical speed-up.

\FloatBarrier
\subsection{Informative Features~$\bs{\gamma}(\bx)$ and Optimal Training Set~$\Ds$}
\label{sec: informative_features}

According to the framework introduced in Section \ref{sec: multi_fidelity_overview}, former versions of \emph{Bayesian Multi-Fidelity Monte-Carlo}~\cite{biehler2015towards,Koutsourelakis2009} have employed~$\ops(\bx)=\ylf(\bx)$, so that model inputs~$\bx$ are entirely filtered through the LF model. This section is devoted to the definition and computation of informative input features~$\bs{\gamma}(\bx)$ that complement the LF model output(s) and represent a key contribution of the paper.

With \emph{informative input feature} we mean in this work a specific path $\gamma(\bx)$ through the input space $\Omega_\bx$. Adding the latter as a further input dimension to the discriminative model (here, a Gaussian process) simplifies the learning of the multi-fidelity conditional in a small data case as the data representation, simply speaking, becomes more structured. An example for an \emph{informative input features} is, for instance, a parameter along which the HF and the LF model behave very differently. To illustrate this effect, we provide snapshots of an extended and simpler multi-fidelity conditional, using one informative feature $\gamma_1$, in Appendix \ref{sec:average_expectation} in Figure \ref{fig:latent_feature}. In the next Section \ref{sec: meaning_features}, we will now provide more insight on how we can efficiently determine such informative features and why incorporating them in the discriminative model is advantageous.

\subsubsection{Informative Features~$\bs{\gamma}(\bx)$ in the Small Data Case}
\label{sec: meaning_features}
The exclusive use of~$\ops(\bx)=\ylf(\bx)$ can lead to a pronounced model error~(\textbf{error 1}), due to the specific choice of discriminative model for the multi-fidelity conditional distribution~$\pdf{\yhf|\ylf}$. While \textbf{error 1} can be mitigated by a more flexible discriminative model, we want to emphasize that this approach is not expedient in the \emph{small data regime}~(small amount of high-fidelity training data) in which we operate.

Instead, we propose to find a \emph{simpler representation} for the multi-fidelity conditional by formulating an extended multi-fidelity conditional in a higher dimensional space. This can be achieved by introducing a specific choice of additional informative input features~$\bs{\gamma}(\bx)$ that come at no extra evaluation costs and complement the LF model output~$\ylf(\bx)$ in the vector~$\ops(\bx)=[\ylf(\bx), \gamma_i(\bx)]^T,\ i\in [0,\nfeature]$. By \emph{simpler representation} we mean that the extended conditional~$\pdf{\yhf|\ops}$ has a lower average variance~(see Appendix \ref{sec:average_expectation})
\begin{equation}
  \label{eqn: average_expectation}
\Ex{\ops}{\Var{\yhf|\ops}{\pdf{\yhf|\ops}}}\le\Ex{\ylf}{\Var{\yhf|\ylf}{\pdf{\yhf|\ylf}}}, 
\end{equation}
such that~(the moments of) the distribution can be approximated with less error for a given number of data points in a \emph{small data regime}~(decrease in the modeling \textbf{error 1}). Additionally, the influence of higher order moments of~$\pdf{\yhf|\ops}$ decreases as well. This decreases the model \textbf{error 1} of a simpler discriminative model, which only resolves the lower order moments of a distribution. On the other hand, the epistemic uncertainty \textbf{error 2} will grow with the dimension of the extended space. These two competing effects are discussed in Section \ref{sec: optimal_num_features}. For now, our goal is to identify a few~$\gamma_i(\bx)$ that, in conjunction with~$\ylf(\bx)$, will reduce as much as possible the model error \textbf{error 1} in \Cref{sec: multi_fidelity_overview} by reducing the average variance~$\Ex{\ops}{\Var{\yhf|\ops}{\pdf{\yhf|\ops}}}$ the most.\\

To simplify the subsequent investigations, we introduce the function~$\dmfx:=\yhf(\bx)-\ylf(\bx)$, respectively the associated random variable~$\dmf:=\yhf-\ylf$. The distribution~$\pdf{\dmf|\ylf}$ is then only shifted but maintains the same conditional variance, such that~$\Var{\yhf|\ylf}{\pdf{\yhf|\ylf}}(\ylf)=\Var{\dmf|\ylf}{\pdf{\dmf|\ylf}}(\ylf)$ holds. We can now formulate the search for optimal~$\gx$ w.r.t. the variance reduction in~$\pdf{\dmf|\ylf}$.

A discriminative model can only exploit the feature~$\gx$~(and therefore reduce the average variance~$\Ex{\ops}{\Var{\dmf|\ops}{\pdf{\dmf|\ops}}}$ effectively) if~$\gamma$ shares a \emph{reasonable smooth and simple}~(low-frequency) trend with~$\dmf$. Otherwise, the discriminative model might not be able to detect a complex structure in~$\pdf{\dmf|\ops}$ and will account functional complexity~(high-frequency changes in~$\pdf{\dmf|\ops}$) as an increase in conditional variance, which would not be target-oriented. 

\begin{figure}[htbp]
 \centering
  \includegraphics[scale=0.21]{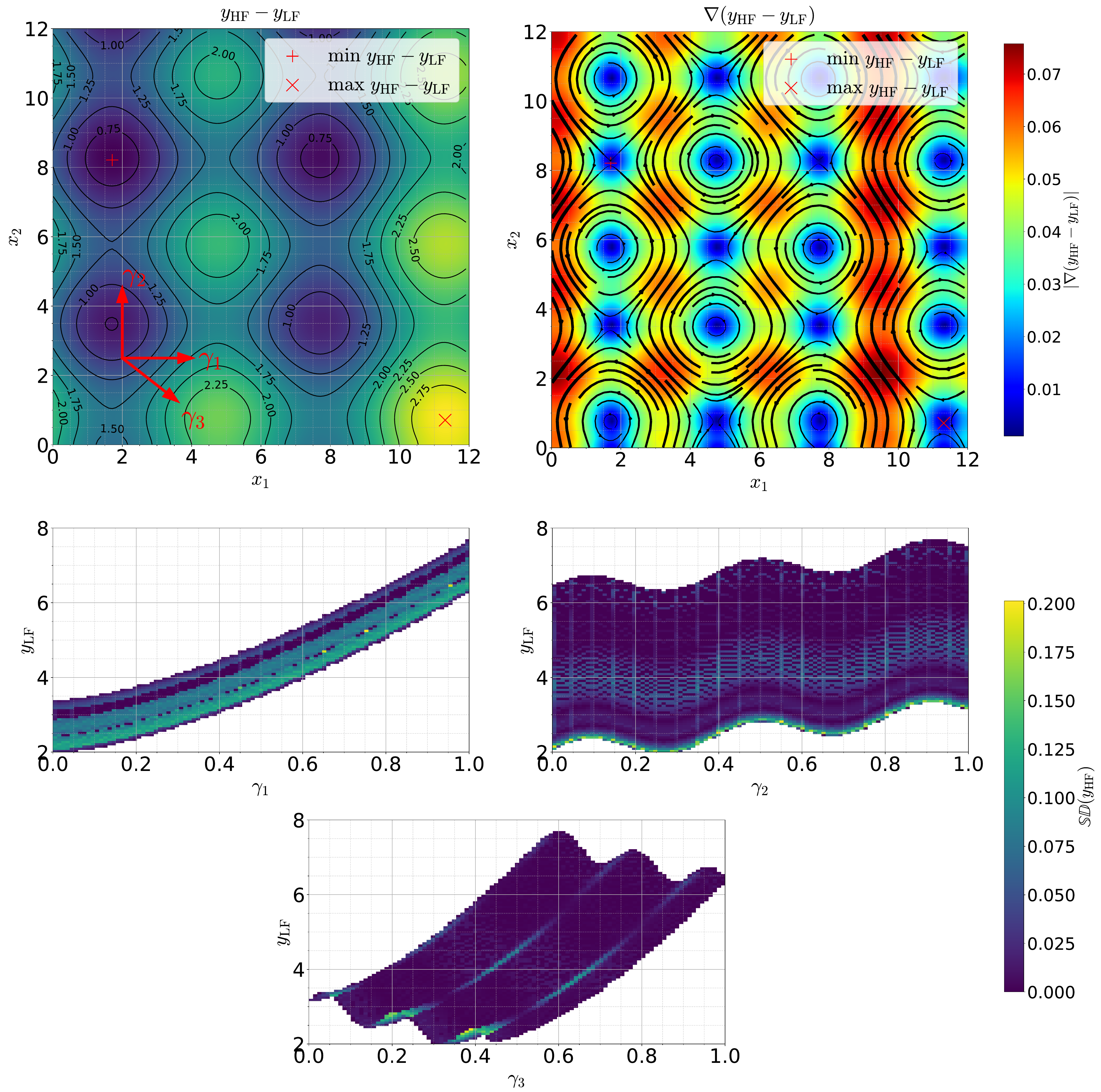}
  \caption{Illustration of different informative input features $\gamma_i(\bx)$: On the top left a difference function ${\dmf(\bx)=\yhf(\bx)-\ylf(\bx)}$ between an HF and a LF model is shown (see Appendix \ref{sec:multi_fid_fun} for the mathematical details). Additionally, we marked the location of its minimum and its maximum. The difference function's gradient is shown as a streamline plot along with the color-coded magnitude of the gradient on the top right. We note that a preference direction with high gradient magnitude can be identified in $x_2$-direction (this becomes specifically clear at the positions $x_1=0$ and $x_1=10$). We use this direction as the informative input feature $\gamma_2(\bx)$. In analogy, the $x_1$-direction is also used as an informative input feature, denoted by $\gamma_1(\bx)$. We furthermore use the direction between the minimum of the difference function and its maximum as an additional informative input feature that is denoted by $\gamma_3(\bx)$. All feature directions are also depicted as red arrows in the top left. In the three bottom plots, we demonstrate the effect of the different informative input features $\gamma_1,\ \gamma_2,\ \gamma_3$ by plotting the features against the LF output $\ylf$ and color-code the standard deviation of the high-fidelity output $\yhf$. It can be seen that $\gamma_2$, which we orientated along the gradient function's preference direction, results in the lowest standard deviation for $\yhf$, such that a model \textbf{error 1} would be reduced most effectively. While $\gamma_3$ shows a similar effect, as it has a significant component in $\gamma_2$-direction, $\gamma_1$ results in the highest standard deviation, having no component in $\gamma_2$-direction.}
  \label{fig: informative_features}
 \end{figure}

To achieve a \emph{reasonable smooth and simple} trend of~$\dmf$ on~$\gx$, we want to find a short and continuous path~$\gx$ between a starting point~$\bxs$ and an end-point~$\bxe$ along which~$\dmf$ experiences additionally a high amount of variance~$\Var{\dmf}{\dmf}\big\vert_{\gx}$, due to changes expressed by the gradient~$\nabla_\bx\dmf$. Without a constraint for the path length, one could imagine a path that repeatably loops in regions of~$\Omega_\bx$ with high variability of~$\dmfx$ or even attempts to cover the entire input space~$\Omega_\bx$. Such a path will however result in a very \emph{complex} mapping~$\dmf(\bx)\big\vert_{\gamma(\bx)}$ along~$\gx$, and therefore also in a complex structure in~$\pdf{\dmf|\ops}$, especially along the additional dimension~$\gamma$~(which is part of the vector~$\ops$). After a suitable path $\gamma$ is identified, the dependency on $\bx$ can be established by, e.g., projecting $\bx$ onto $\gamma$. For the sake of simplicity, we investigate informative features on straight lines in the following, simplifying the projection step, as well. Figure \ref{fig: informative_features} visualizes the effect of different feature directions reusing the exemplary multi-fidelity problem from Figure \ref{fig: conditionals}. The Figure illustrates how a direction that follows the maximum gradient $\nabla\dmf$ (on average), most effectively reduces the variance, respectively the standard deviation, in the $\yhf$-dimension, reducing the model \textbf{error 1}.

The use of the difference function's gradient~$\nabla_\bx \dmfx=\nabla_\bx\yhf(\bx) - \nabla_\bx\ylf(\bx)$ to determine the path~$\gx$ is usually prohibitive due to the cost of HF model evaluations in combination with a general lack of gradient~(or adjoint formulations) for the model output w.r.t. the input space, for most problems of practical interest relying on legacy codes. A more in-depth analysis of efficient approaches to determine paths~$\gx$ that (approximately) incorporate the gradient information~$\nabla_\bx\dmfx$ and fulfill the complexity constraints is outside the scope of this paper. 

\begin{remark}[Generalization of and comparison to other existing multi-fidelity frameworks]
We note that employing the whole input vector~$\ops(\bx)=[\ylf(\bx),\bx]^T$~(meaning~$\bs{\gamma}(\bx)\equiv \bx$) as, e.g., in~\cite{perdikaris2017nonlinear} for \emph{nonlinear autoregressive multi-fidelity GP regression}~(\emph{NARGP}), would only be applicable to low-dimensional~$\bx$, due to the curse of dimensionality and the resulting accuracy issues for surrogates in high-dimensional spaces in combination with very limited training data. 
Our intention is not to develop a new surrogate approach, such as \emph{NARGP}, but rather to propose a new sampling strategy, which keeps the advantageous property of standard Monte-Carlo techniques of being error-independent of the underlying problem's stochastic dimension. In the case of \emph{BMFMC} we correct the sampling on the LF model by exploiting a \emph{low dimensional} statistical relationship between the fidelity levels and a small set of informative input features. The proposed method has its strength in \emph{high-dimensional} stochastic settings. For lower dimensional UQ problems, direct multi-fidelity surrogate approaches, such as \emph{NARGP}, might be preferable if the necessary training data can be afforded. For dynamic problems, it was shown in~\cite{lee2018linking} that it is advantageous to incorporate time~$t$ and time derivatives in the form of time shifts~$\ylf(\bx,t),\ \ylf(\bx,t+\Delta t)$, of the LF simulation outputs as further features. This idea can directly be integrated in our approach, but without the necessity to treat~$\bx$ explicitly, by choosing, e.g.,~${\ops(\bx,t)=\left[\ylf(\bx,t),t,\ylf(\bx,t+\Delta t)\right]^T}$.
\end{remark}

\begin{remark}[Multiple low-fidelity models]
The \BMFMC framework also allows the incorporation of several low-fidelity models. There are two main options for introducing another LF model into the framework. We postpone investigations concerning the distribution of the computational budget for multiple LF models to our future research. In a \emph{parallel incorporation}, one or multiple further LF models are  added to the low-fidelity feature vector~$\ops(\bx)=[\ylf_1(\bx), \ylf_2(\bx),\dots, \gamma_i(\bx)]^T$. In this scenario, sampling over all LF models is required, which adds to the overall cost of the method. The \emph{parallel incorporation} strategy might, however, pay off if each of the additional LF models furnishes exclusive information about the HF model, and hence also reduce the average variance~$\Ex{\ops}{\Var{\dmf|\ops}{\pdf{\dmf|\ops}}}$, effectively. 
Alternatively, a \emph{sequential incorporation} of LF models is advantageous if the LF models have different accuracy and cost levels. Therefore, a hierarchical structure of model fidelities can be established. For two LF models, LF 1 and LF 2, this idea can be formulated with a chained multi-fidelity conditional~$\pdf{\yhf|\ops_2}=\int\pdf{\yhf|\ops_1} \cdot \pdf{\ops_1|\ops_2}d\ops_1$ and only requires sampling over LF 2. The two multi-fidelity conditionals~$\pdf{\yhf|\ops_1}$ and~$\pdf{\ops_1|\ops_2}$ can be learned in a small data regime, and integration is performed through Monte-Carlo sampling on LF 1. Such a formulation can be beneficial if the variance in the two conditionals is low compared to the conditional variance when~$\pdf{\yhf|\ops_2}$ is approximated directly.
\end{remark}

We now present a simple and robust heuristic that incorporates many of the features described for an optimal feature~$\gx$ but comes at no additional computational cost. We propose to choose~$\gamma_i(\bx)$ from a lower-dimensional representation~$\xred$ of the original input~$\bx$~(see Appendix \ref{sec:dim_reduction}). Specifically, we select the~$\xred_i$ as~$\gamma_i(\bx)$, which cause the highest output variance~$\Var{\ylf}{\pdf{\ylf}}$ on the LF model along the direction~$\xred_i$. This means that a small step in the direction~$\Delta \xred_i$ has on average also a high~(directional) variance gradient~$\xred_i\cdot\nabla_\bx\Var{\yhf}{\pdf{\yhf}}$. The heuristic assumes that the difference in the LF and HF gradient and hence the gradient of the difference function~$\nabla_\bx\dmfx$ is high when the magnitude of the~(directional) gradient of the LF model~$\xred_i\cdot\nabla_\bx\ylf(\bx)$ is high. This is usually a good guess for most problems and has the advantage that, due to the sampling on the LF model, enough data is available to determine appropriate directions with confidence. 
To determine suitable candidates for~$\gamma_i(\bx)$, we define a correlation measure~$\mathbf{r}$ between the individual dimensions of the reduced input vector~$\xred$ and the LF simulation output~$\ylf$, using the projection of the corresponding reduced input matrix~$\Xred$ on the LF output vector~$\Ytestlf$:
\begin{equation}
 \mathbf{r}={\big{\vert}}{{\Xred}^{T}\cdot {{Y}_{\text{LF}}}^{*}}{\big{\vert}} \label{eqn: LF_sensitivityb}
\end{equation} 
We furthermore note that a direction along a~(reduced) input space dimension automatically fulfills the complexity constraint for~$\gamma$ and has the form~$\gamma(\bx)=\xred_i$. Here,~$\xred = C(\bx)$ with~$C$ being the compression operation on the original input~$\bx$ and~$\xred_i$ being a particular dimension of the reduced input space. The definition~\eqref{eqn: LF_sensitivityb} is the absolute value of the scaled Pearson correlation coefficient, calculated for each dimension of~$\xred$ and the LF output~$\ylf$. According to the previously described assumptions, input dimensions~${\hat{x}}_j$ that show high values for~$r_j$ will cause the most variance~$\Var{\ylf}{\ylf}$ among the considered directions and lead to an efficient reduction of the expected variance in Equation~\eqref{eqn: average_expectation}.

We select the~${\hat{x}}_i$ that correspond to the~$i$-highest entries~$r_i$ in~$\bs{r}$, with~$i\in\mathbb{N}:[1,\mathrm{n}_{\bs{\gamma}}]$, as an informative feature~$\gamma_i$ of the input. In our numerical examples~(see Section \ref{sec: numerical_demonstration}) and in the small data regime~(up to 300 HF and 300 LF simulation runs for the construction of the probabilistic multi-fidelity regression model), one or two additional LF features~$\gamma_1$ and~$\gamma_2$ gave the best results. More elaboration on this aspect is provided in the next Section \ref{sec: optimal_num_features}.\\

\begin{remark}[Nonlinear correlation measure]
  As an alternative to the proposed linear correlation measure in Equation~\eqref{eqn: LF_sensitivityb} the procedure can be easily adapted to a nonlinear correlation measure by using a nonlinear kernel function~$k(\bx,\bs{y})$ that returns the correlation matrix~$K = k(\Xred, \Ytestlf)$:
  \begin{align}
    \begin{split}
   \mathbf{r}_{\text{nl}}&={\big{\vert}}{{\Xred}^{T}\cdot K\cdot {{Y}_{\text{LF}}}^{*}}{\big{\vert}}
    \end{split}
  \end{align} 
  A possible choice for such a kernel function could be the squared exponential kernel~(see Equation~\eqref{eqn: s_exp_kernel}). However, we leave the investigation of such a correlation measure to our future research.
  \end{remark}

\begin{remark}[Latent variable model: Marginalization of the informative features~$\bs{\gamma}$]
  An interesting viewpoint on the effect of the additional informative features is the interpretation as a latent variable model~\cite{bishop1998latent}. The idea of latent variables is a very powerful and well-known framework in~(probabilistic) machine learning.  Here, auxiliary or latent variables are introduced that define a simpler joint distribution with the original variables. The latter can then be integrated~(marginalized) over the latent variables to yield the actual distribution of interest.
  
  In our application, the informative features~$\bs{\gamma}$ play the role of latent variables. Reformulating the multi-fidelity uncertainty quantification problem given in Equation~\eqref{eqn: uq_extended} shows that the extended multi-fidelity conditional distribution~$\pdf{\yhf|\ops}$ is actually a latent variable model for~$\pdf{\yhf|\ylf}$~(see original Equation~\eqref{eqn: mf_uq}):
  \begin{align}
    \label{eqn:latent_marginalization}
    \begin{split}
    \pdf{\yhf|\ylf}&=\int\limits_{\ssp{\bs{\gamma}}} \underbrace{\pdf{\yhf | \ylf, \bs{\gamma}}}_{\pdf{\yhf|\ops}}\cdot \underbrace{\pdf{\bs{\gamma}|\ylf}}_{\pdf{\ops|\ylf}}~ d \bs{\gamma}\\
    &\approx \frac{1}{\nsample}\sum\limits_{j=1}^{\nsample} \pdf{\yhf|\ylf, \bs{\gamma}_{j}}, \quad\text{with } \bs{\gamma}_{j} \sim \pdf{\bs{\gamma}|\ylf}
    \end{split}
  \end{align}
We select $\bs{\gamma}$ such that the average variance of $\pdf{\yhf|\ops}$ is lower than in $\pdf{\yhf|\ylf}$. The extended density $\pdf{\yhf|\ops}$ can hence be better approximated in the \emph{small data regime}, e.g., by a Gaussian Process, than~$\pdf{\yhf|\ylf}$. Furthermore, sampling from~$\pdf{\bs{\gamma}_{i}|\ylf}$ is trivial as the joint distribution~$\pdf{\ylf,\bs{\gamma}}$ is already available in the form of discrete samples due to the prior sampling on the LF model.
  
  Effectively, Equation~\eqref{eqn:latent_marginalization} yields a \emph{conditionally non-Gaussian approximation} of~$\pdf{\yhf|\ylf}$ due to the marginalization, which lowers the \textbf{model error 1}, drastically. Please note that~$\pdf{\yhf | \ylf, \bs{\gamma}}$ is still approximated by a Gaussian Process but the additional structure, introduced by~$\bs{\gamma}$, as well as the generally non-Gaussian distribution~$\pdf{\bs{\gamma}|\ylf}$, render the resulting marginal distribution~$\pdf{\yhf|\ylf}$ non-Gaussian. 
  In the \emph{small data regime} even a more flexible discriminative model would not be able to find a better approximation to~$\pdf{\yhf|\ylf}$ when applied directly to the data-set~$\{\Ytrainhf, \Ytrainlf\}$, while potentially even raising regularization issues. In contrast, the latent variable approach allows simpler and more robust discriminative models, such as Gaussian Processes. Additionally, it increases controllability as the informative features~$\bs{\gamma}$ can be selected based on predefined criteria, as described later in this section.
  \end{remark}
\FloatBarrier
\subsubsection{Number of Informative Features: Model-Error Versus Epistemic Uncertainty}
\label{sec: optimal_num_features}
\begin{figure}[!htb]
 \centering
  \includegraphics[scale=0.27]{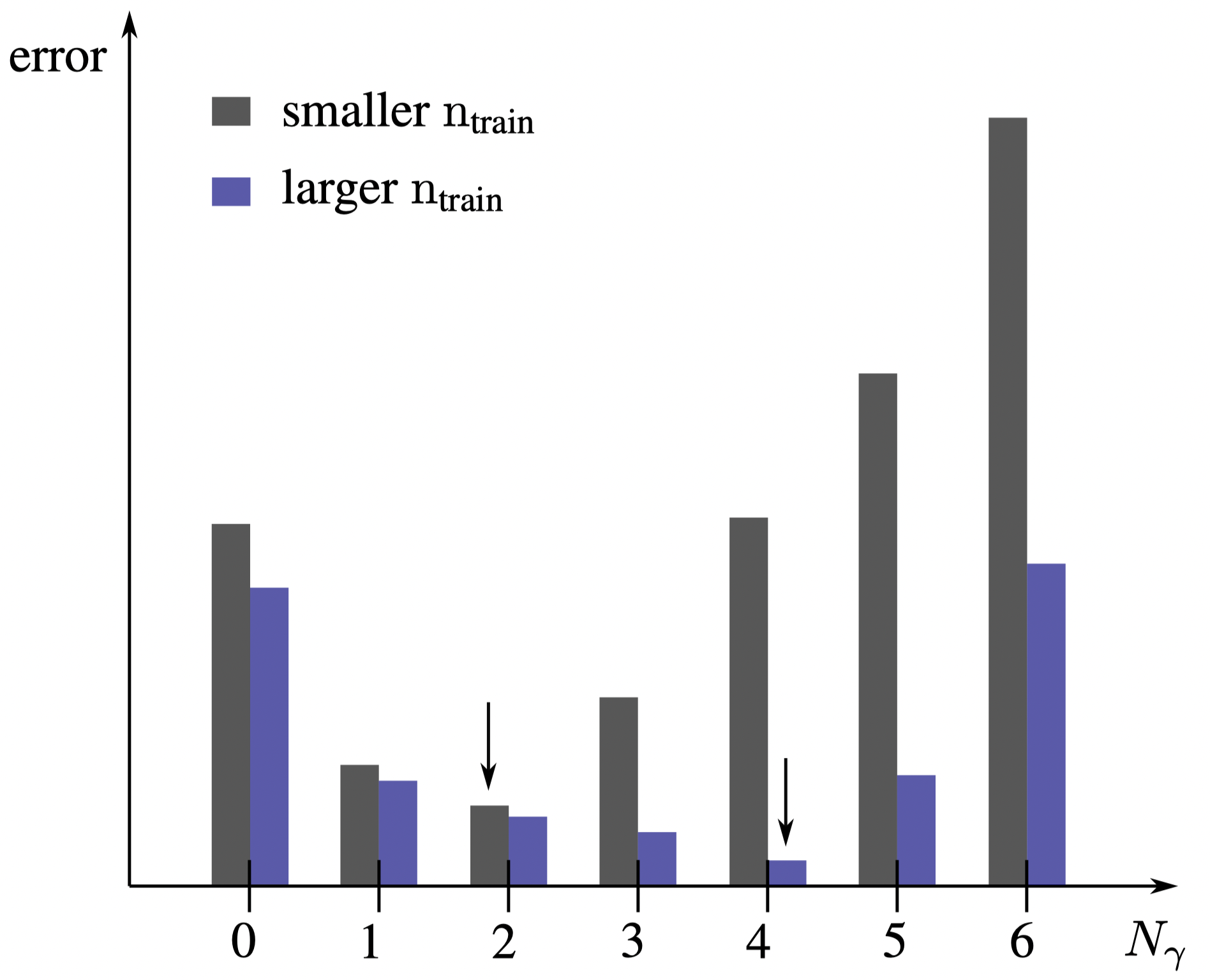}
  \caption{Schematic illustration of the error behavior in~$\pdf{\yhf|\ops}$ for an increasing number of features,~$\gamma_i$. The incorporation of features leads at first to a decrease of the modeling \textbf{error 1}, before a larger~$\Omega_{\ops}$ leads again to an increase of~(\textbf{error 2}). For a larger training data size~$\ntrain$ the error is in general lower and the minimum error~(indicated by arrows) lies at a higher number of features. See \Cref{fig: reduction_features_cylinder} for a numerical demonstration.}
  \label{fig: error_effects}
 \end{figure}

Suppose too many informative features~$\gamma_i$ are added to~$\ops$. In that case, the resulting space~$\ssp{\ops}$ becomes too large to be sufficiently covered by~$\Ds$ in a \emph{small data regime} so that \textbf{error 2} will increase due to growing \emph{epistemic uncertainty}. \Cref{fig: error_effects} schematically depicts this effect. If it is possible to realize a larger amount of training data, more informative features~$\gamma_i$ can be added to decrease the approximation error further.
 As depicted in Figure \ref{fig: error_effects} this leads to an error minimum over the number of employed informative features. The minimum location depends on the number of training data points~$\ntrain$ and moves to a larger number of informative features for increasing~$\ntrain$. In all our investigations, where we used up to~$\ntrain=300$ HF and LF simulations runs to approximate the probabilistic multi-fidelity conditional~$\pdf{\yhf|\ops}$, one or two informative features usually gave the best results. This is equal to a space coverage of roughly ten training points per dimension if all points are organized on a uniform grid. The same grid density for only one more informative feature~(three informative features plus the LF model dimension) requires 10000 HF-LF model runs for~$\Ds$, which is usually beyond the affordable computational budget for the applications we are interested in.

\FloatBarrier
\subsubsection{Optimal Selection of Training Points~$\Ds$}
\label{sec: training_data}
For an optimal selection of training points~$\Ds$, given a computational budget of~$\ntrain$ HF model runs, we propose the following procedure:
Given the large LF data set~${\Dslfx\coloneqq\{\Xtest,\Ytestlf\}}$, from the sampling of the LF model, we select the training inputs~$\Xtrain$ to be a specific space-filling subset of the input samples~$\Xtest$, such that~$\Xtrain\subset\Xtest$, as we have already run the LF model simulations for these inputs. We elaborate on the space-filling properties in the sequel. We then run the associated HF simulations according to~${\Ytrainhf=\yhf(\Xtrain)}$. The training data set is then defined as~$\Ds=\{\Ytrainhf,\Opstrain\}$. 

Usually, the initial training data selection for surrogate models of computer experiments aims for a space-filling design strategy in the entire input space~$\Omega_\bx$, i.e., a Latin Hyper-Cube design to explore the input space efficiently~\cite{kang2015system,jones2009design,ranjan2011computationally} or quasi Monte-Carlo~(QMC) sequences~\cite{caflisch1998monte, sobol1990quasi}. As the input space is assumed to be large, it is advantageous to demand \emph{space-filling properties} only w.r.t. the important part of~$\ssp{\bx}$, represented by~$\featExt=\gamma_i$, with~$i\in\mathbb{N}:[1,\mathrm{N}_{\featExt}]$ and~$\mathrm{N}_{\featExt}$ being the number of input features used in the extended vector~$\featExt$. We note that~$\mathrm{N}_{\featExt}>\mathrm{N}_{\bs{\gamma}}$, such that we demand space-filling properties to more dimensions~$\gamma_i$ than used to define~$\ops$. This is necessary to yield a training design that captures the uncertainty of~$\pdf{\yhf|\ops}$ as well. The next informative features that were not used in~$\ops$ are believed to have the most impact on the conditional variance of~$\pdf{\yhf|\ops}$~(see discussion in Section \ref{sec: meaning_features}).

Given the training data size~$\ntrain$, we choose a space-filling subset~$\FeattrainExt\subset\FeattestExt$ from the large sampling data set size~$\nsample$, with~$\ntrain\ll\nsample$. In the numerical implementation, we used a \emph{diverse-subset algorithm} based on Wessing and Salomon~\cite{wessing2015two,salomon2013psa} to reuse the already computed LF model runs.

Given~$\OpstrainExt$ and the corresponding~$\Xtrain$, we can now run the HF simulations accordingly to~$\Ytrainhf=\yhf(\Xtrain)$. We found that~${\mathrm{n}_{\featExt}\in\mathbb{N}:[3,6]}$ is a good choice in the small data regime for around~$\ntrain=300$ training points. 

\FloatBarrier
\subsection{Summary of the BMFMC-Algorithm}
\label{sec: algorithmic_summary}
We can summarize the necessary steps for Bayesian multi-fidelity Monte-Carlo in Algorithm~\ref{alg: pseudo-code}. The sub-algorithms in Algorithm \ref{alg: trainingdata} and Algorithm \ref{alg: posteriorstatistics} in Appendix \ref{sec: sub_algorithms} show the details of the determination of~$\ops$, discussed in the previous section and the implementation of the posterior statistics from Equation~\eqref{eqn: cond_expectation}-\eqref{eqn: cond_variance}.

\begin{algorithm}
\caption{Pseudo-code for Bayesian multi-fidelity Monte-Carlo~(BMFMC)}\label{alg: pseudo-code}
\begin{algorithmic}[1]
\Require~$\pdf{\bx},\ \yhf(\bx),\ \ylf(\bx),\ \ntrain, \ \nsample,\ \bs{y}_{\text{HF,support}}$
\State~$\Xtest$ = \Call{Generate}{$\pdf{\bx},\nsample$}\Comment{Draw~$\nsample$ samples from~$\pdf{\bx}$}
\State~$\Ytestlf\gets\ylf(\Xtest)$\Comment{Run LF model for~$\nsample$ samples}
\State~$\pdf{\fmod|\ops}$ = \Call{DesignPriorGP}{$\Ytestlf$}
\State~$\Ds$ = \Call{TrainData}{$\Xtest,\Ytestlf,\ntrain,\nsample$}\Comment{Here the HF model is evaluated $\ntrain$ times}
\State~$\mathcal{GP}_{\fmod|\Ds}\gets\mathcal{GP}_{\fmod}$\Comment{Train GP model on~$\Ds$}
\item[]
\State \Return~$\bs{p}_{\yhf,\mathbb{E}^*}$,~$\bs{p}_{\yhf,\mathbb{V}^*}$ = \Call{PosteriorStatistics}{$\mathcal{GP}_{\fmod|\Ds},\Opstest,\nsample$}
\end{algorithmic}
\end{algorithm}
At first, we sample~$\nsample$ samples from the input density~$\pdf{\bx}$ and store them in the matrix~$\Xtest$. These samples are then run on the LF model, which yields the LF output matrix~$\Ytestlf$. 
Afterwards, we set up the prior Gaussian Process that is used to approximate the multi-fidelity conditional~$\pdf{\yhf|\ops}$. For this step we determine the informative features~$\gamma_i(\bx)$ as described in Section \ref{sec: meaning_features} and construct the low-fidelity feature vector~$\ops=[\ylf,\gamma_i]^T,\ i\in[0,\nfeature]$. 
Subsequently, we select an appropriate set of training data~$\Ds$~(see Algorithm \ref{alg: trainingdata} in Appendix \ref{sec: sub_algorithms} for details) following the space-strategy that was described in Section \ref{sec: training_data}.
Eventually we can then generate the posterior statistics from Equations~\eqref{eqn: cond_expectation} and~\eqref{eqn: cond_variance} using the Algorithm described in Appendix \ref{sec: sub_algorithms}.

\FloatBarrier
\subsection{Considerations of Computational Complexity and Numerical Speed-up}
\label{sec: aspects_implementation}
The overall costs for \emph{BMFMC} are composed of the sampling costs for the LF model and the costs for the HF and LF simulations in the training data~$\Ds$. We denote the average speed-up factor between an HF and a LF simulation run by~$f_{\text{HF/LF}}:=\frac{\text{costs HF}}{\text{costs LF}}$. LF models can be motivated by pure numerical relaxation, simplified physics, geometry, or combinations of these aspects. A more detailed discussion of computational complexity, simulation costs, and the theoretical speed-up for numerical relaxation can be found in Appendix \ref{appendix: comp_cost}. The speed-up through \emph{BMFMC} compared to the computational costs of a classic~(Monte-Carlo) sampling strategy is then given by:
\begin{align}
    \label{eqn: speed-up}
    \text{speed-up \BMFMC}&\coloneqq \frac{\mathrm{N}_{\text{MC,HF}}\cdot \text{costs HF}}{\mathrm{N}_{\text{MC,LF}}\cdot \text{costs LF} + \ntrain\cdot \text{costs HF}}=\frac{\mathrm{N}_{\text{MC,HF}}\cdot f_{\text{HF/LF}}}{\mathrm{N}_{\text{MC,LF}} + \ntrain\cdot f_{\text{HF/LF}}}
\end{align}

\begin{figure}[htbp]
 \centering
  \includegraphics[scale=0.25]{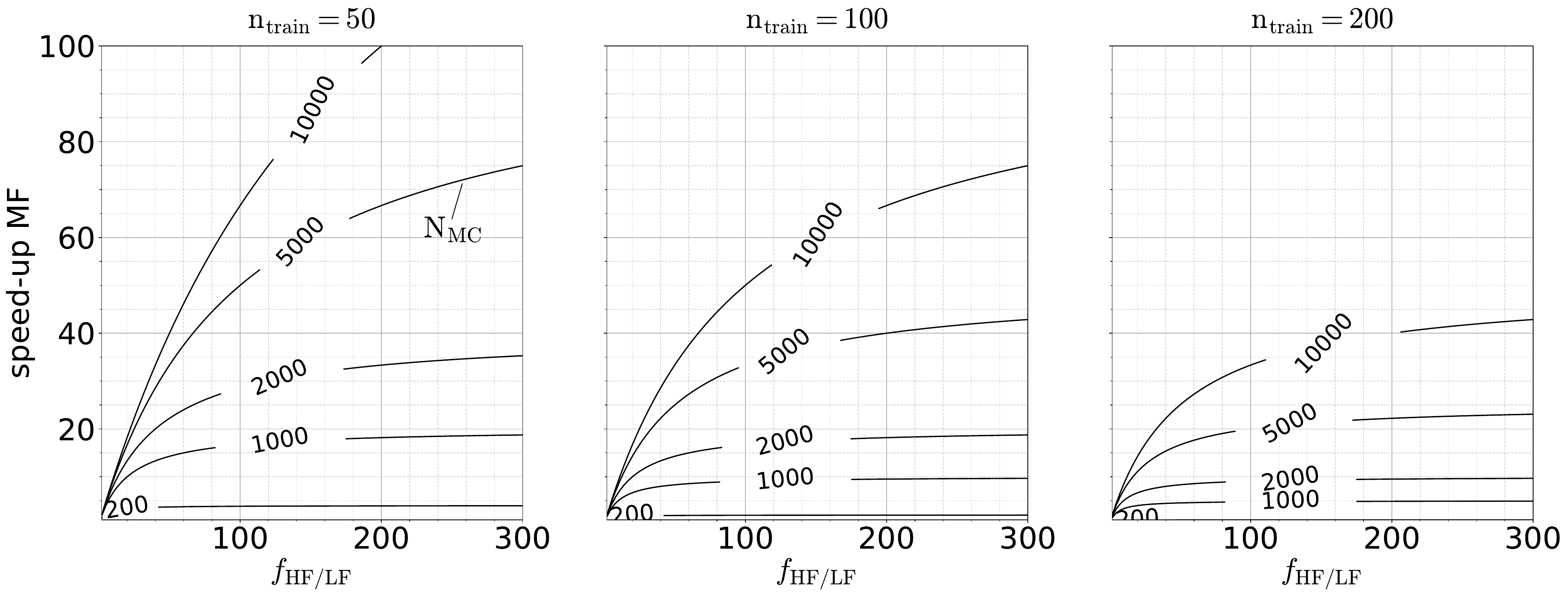}
  \caption{Overall speed-up of the proposed multi-fidelity Monte-Carlo~(\emph{BMFMC}) approach for UQ for different training data sizes~$\ntrain$ and Monte-Carlo sample sizes~$\mathrm{N}_{\text{MC}}$ as well as different HF/LF model speed-ups.} 
  \label{fig: speed_up_mf}
 \end{figure}

\Cref{fig: speed_up_mf} shows the theoretical speed-ups of \BMFMC for different LF model speed-ups~$f_{\text{HF/LF}}$, as well as different training data sizes~$\ntrain$ and Monte-Carlo sample sizes~$\mathrm{N}_{\text{MC}}$.
 The multi-fidelity approach becomes especially powerful if a high number of Monte-Carlo evaluations on the HF model would have been necessary to estimate a statistic of interest with high accuracy. 

\FloatBarrier
\section{Numerical Demonstration}
\label{sec: numerical_demonstration}
In the following numerical examples, we demonstrate the accuracy and efficiency of the proposed generalized Bayesian multi-fidelity Monte-Carlo framework. The LF models deployed are automatically generated by numerical relaxation of the corresponding HF model, as described in \Cref{sec: aspects_implementation}. We implemented the generalized multi-fidelity approach for uncertainty quantification \emph{BMFMC} in \emph{QUEENS}~(Quantification of Uncertainties in Engineering and Science)~\cite{queens}, a software platform for uncertainty quantification, physics-informed machine learning, Bayesian optimization, inverse problems, and simulation analytics. \emph{QUEENS} is capable of interacting with a variety of commercial, open-source, and in-house simulation engines and enables the fully automatic set-up of all required simulations on high-performance computing~(HPC) clusters, workstations, and desktop computers. We solved the stochastic flow past a cylinder problem on a workstation with Intel Core i7-8000K CPUs running at 3.7 GHz. The second demonstration of a stochastic fluid-structure interaction problem was computed on an HPC cluster with Intel Xeon E5-2680v3 "Haswell" CPU running at 2.5 GHz.

\subsection{Stochastic Flow Past a Cylinder: High-Order Discontinuous Galerkin Navier-Stokes Solver}
For the first numerical demonstration, we investigate uncertainty propagation for a widely used benchmark in computational fluid dynamics: The flow past cylinder test case for incompressible flows, as defined by Sch\"afer and Turek~\cite{schafer1996benchmark}. Similar setups have also been discussed (see Perdikaris et al.~\cite{perdikaris2015multi}). The geometry of the two-dimensional domain is a rectangular channel with height~$H=0.41$ and length~$L=2.2$, as depicted in \Cref{fig: cylinder_problem}. We modify the original benchmark problem at~$Re=100$ to a stochastic flow problem for our investigations on efficient uncertainty propagation:
\begin{figure}[htbp]
 \centering
  \includegraphics[scale=3.3]{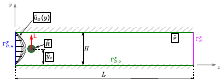}
  \caption{Setup of the stochastic flow past a cylinder problem. Random inputs are written in boxes and have a tilde superscript.} 
  \label{fig: cylinder_problem}
 \end{figure}
A circular cylinder with uncertain radius~$\tilde{R}~\sim~\ud{R}{0.035,0.07}$ is placed in the channel at position~$x_c=0.2$ in streamwise direction. The cylinder's distance to the bottom channel wall is also a univariate and uniform random variable~$\tilde{y}_c~\sim~\ud{y}{0.16,0.24}$. No-slip boundary conditions are imposed on~$\Gamma_{D,0}^{\mathcal{F}}$, defined by the cylinder surface and the channel walls~(marked in green in \Cref{fig: cylinder_problem}). At the outflow boundary~$\Gamma_{N}^{\mathcal{F}}$~(shown in magenta), a Neumann boundary condition is prescribed as described in~\cite{fehn2017stability}. Additionally, the kinematic viscosity~$\tilde{\nu}$ of the fluid is uncertain and modeled as a random variable with the uniform distribution~$\ud{\nu}{9.5\cdot10^{-4},1.5\cdot 10^{-3}}$. Furthermore, we assume a transient and stochastic Dirichlet boundary condition on the inflow section~$\Gamma_{D,u}^{\mathcal{F}}$~(shown in blue) in the form of a random field in space with a sinusoidal ramp over time:
\begin{align}
\label{eqn: Dirichlet_BC}
\begin{split}
\Gamma_{D,u}^{\mathcal{F}}: \quad \mathbf{u} &=
\begin{bmatrix}
\tilde{u}_x(y,t)\\
0\\
0
\end{bmatrix},\\
\tilde{u}_x(y,t) & =\bigg[\underbrace{U_{\rm{m}} \frac{4 y (H-y)}{H^2}}_{\text{mean function for }t=T/2}+\underbrace{\GP{}{0,\tilde{k}(y,{y}')}}_{\text{Non-stat. random process}}\bigg]\cdot\underbrace{\sin(\pi t/T)}_{\text{transient ramping}}
\end{split}
\end{align}
The quantity of interest is the maximum lift coefficient~$C_{L,\max}$, due to the lift force~$L$ on the cylinder (shown in red) in the y-direction. In the uncertainty propagation problem, we want to infer the distribution~$\pdf{C_{L,\max}}$ as a stochastic response to the uncertain boundary condition and parameters, whose distribution we abbreviate by~$\pdf{\bx}$.
The random process is modeled as a Gaussian Process with a non-stationary kernel function~$\tilde{k}(y,{y}')$, so that the standard deviation of the process is~$\frac{1}{8}$ of the mean inflow~$\mu_u(y,t)$. The non-stationary covariance function is formulated as a stationary squared exponential covariance function~$\kfun{u}{y}{y}$ with space-dependent signal-variance~$\left(\sigma_u(y)\right)^2$:
\begin{equation}
\label{eqn: custom_kernel}
\tilde{k}(y,{y}')=\underbrace{\left(0.125\cdot\mu_u(y)\right)^2}_{\left(\sigma_u(y)\right)^2}
 \cdot \underbrace{\exp \left(-\frac{(y-{y}')^2}{2\ls^2}\right)}_{\kfun{u}{y}{y}}.
\end{equation}
Details concerning the numerical generation of random samples, as well as an overview of the stochastic model setup, can be found in Appendix \ref{sec: details_cylinder}.

The random input~$\bx\sim \pdf{\bx}$ is simulated over a time interval of~$0 \leq t \leq T=8$, with a zero velocity field in the domain at the initial time~$t=0$. We solve the uncertainty propagation problem using two fidelity levels of a high-order discontinuous Galerkin~($L^2$-conforming) discretization developed in~\cite{fehn2017stability,fehn2018turbulence}. 
From a practical perspective, the flexibility to vary the polynomial degree~$k$ of the shape functions and the mesh resolution~$h$ independently, as a means to increase the spatial approximation properties of the discretization, is attractive as one does not have to generate several meshes. We exploit this property for the multi-fidelity approach and define a high-fidelity model version with polynomial degree~$k=6$ and a low-fidelity version of the benchmark using~$k=3$ to approximate the velocity field.

\begin{figure}[htbp]
 \centering
  \includegraphics[scale=1.0]{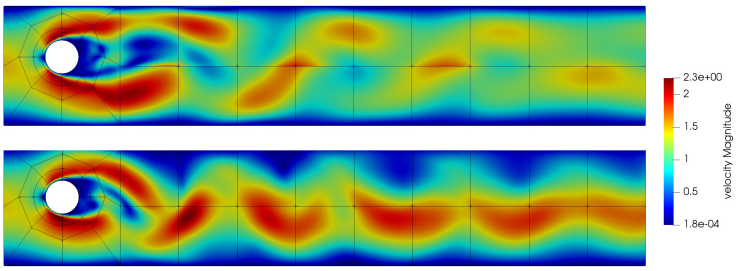}
  \caption{Example snapshot of the velocity magnitude for a low-fidelity simulation with~$k=3$~(top) and high-fidelity model simulation with~$k=6$~(bottom). Both simulations used identical inputs~$\bx$ and are shown for the same simulation time~$t=T/2$.} 
  \label{fig: cylinder_sim}
 \end{figure}

The simulation domain and the mesh for one random input realization~$\bx\sim\pdf{\bx}$ are shown in \Cref{fig: cylinder_sim} for the HF and the LF model version, respectively. The reduction of the polynomial degree from~$k=6$ to~$k=3$ leads to a speed-up of roughly eight, which agrees with Equation~\eqref{eqn: speedup}~(see Appendix \ref{appendix: comp_cost}). The numerical example illustrates a simple way to generate LF models by numerical relaxation without the intention to show maximal possible speed-up of the method. Greater speed-ups can be achieved by combining numerical relaxation, geometric representation, and simplified physics (see, e.g., \cite{biehler2015towards}).

The procedure for the multi-fidelity uncertainty propagation follows algorithm \ref{alg: pseudo-code}. We run simulations for the~${\nsample=10000}$ input realizations of~$\pdf{\bx}$ stored in~$\Xtrain$, with~$\bx=\left[\tilde{u}_x(y),\tilde{\nu},\tilde{R},\tilde{y}_c\right]^T$ on the LF version of the cylinder flow problem and obtain a vector~$\Ytestlf$ of according LF model responses for~$C_{L,\max}$.

\begin{remark}[Number of sample points]
The necessary number of sample points for the LF model is problem-dependent and strongly relies on the goal of the analysis. As we are interested in the entire density, the amount of necessary sample points for an accurate approximation is significantly higher than for point estimates, such as the mean value or the variance. The convergence of the density estimate or statistic of interest can be investigated over an increasing number of sample points. Error \emph{estimates} exist only for point estimators such as Monte-Carlo mean or variance estimators but can be used for an initial orientation of the sample size~$\nsample$~\cite{wan2014estimating}. The standard error of the Monte-Carlo estimator for the mean yields:
\begin{align}
    \label{rsme_mc}
    \sigma_{\mathbb{E}}
    \approx\frac{\hat{\sigma}}{\sqrt{\nsample}},
\end{align}
where~$\sigma_{\mathbb{E}}$ is the standard deviation of Monte-Carlo error,~$\hat{\sigma}$ is the estimate of the standard deviation of the QoI and~$\nsample$ is the number of sample points. We select an initial sample size of~$\nsample=10000$ for the LF model, so that the relative error~$\sigma_{\mathbb{E}}/\hat{\sigma}$ in the mean estimate is 1 \%.
\end{remark}

\begin{figure}[htbp]
 \centering
  \includegraphics[scale=0.25]{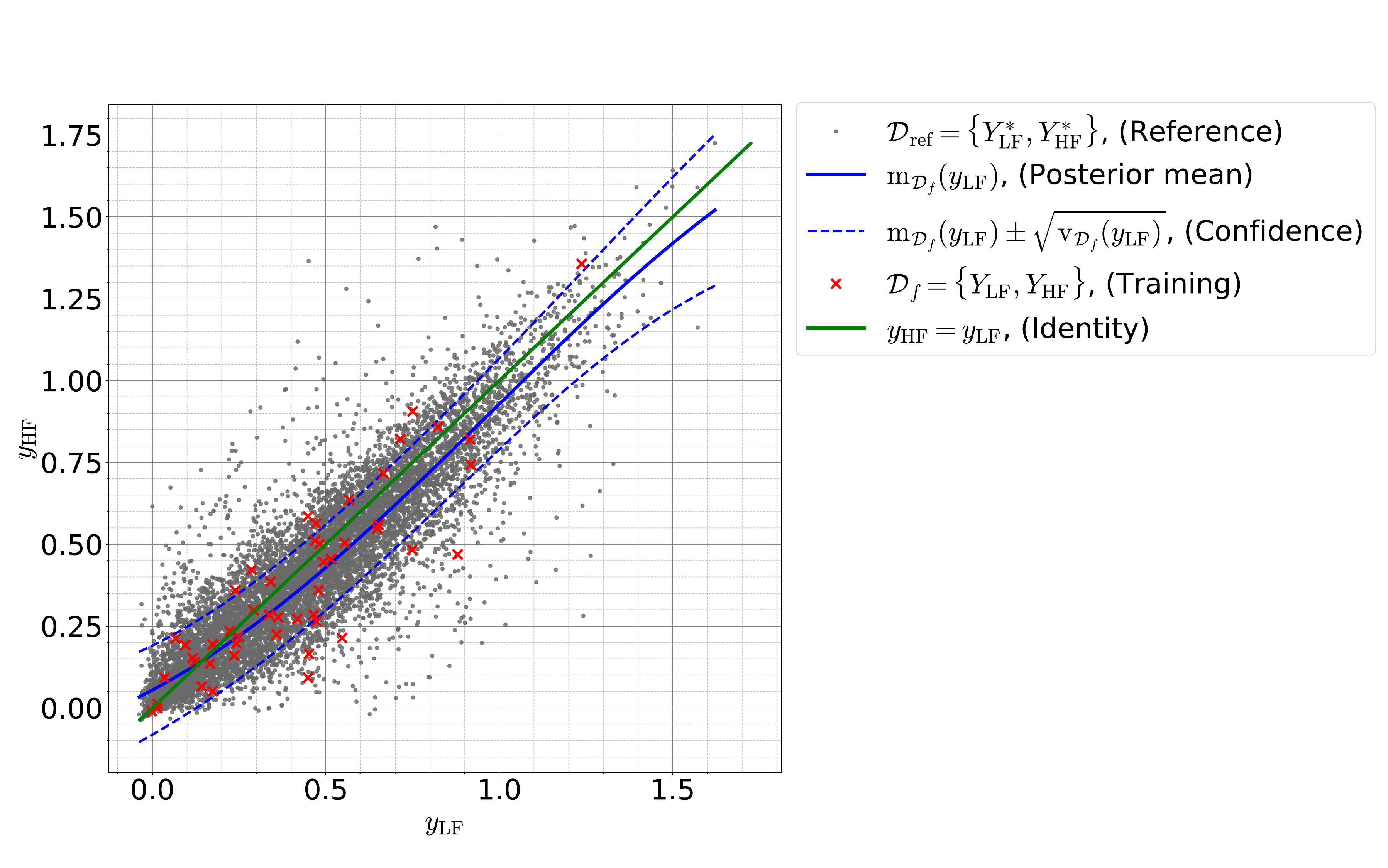}
  \caption{LF and HF output tuples of the flow past a cylinder problem. The posterior Gaussian Process for the approximation of~$\pdf{\ftest|\Ds}$ is shown in form of its mean function~$\mf{\ylf}$ and associated credible intervals. The training data~$\Ds$ is marked by red crosses.} 
  \label{fig: ylf_yhf_cylinder}
 \end{figure}

Afterwards, we successively compute features~$\gamma_i$ and choose five features to calculate an~$\ssp{\gamma_i\times\ylf}$-filling subset~$[\ylf(\Xtrain),\gamma_i(\Xtrain)]^T\subset[\ylf(\Xtest),\gamma_i(\Xtest)]^T$, with~$i\in\mathbb{N}:[1,5]$. We choose a data set of size~$\ntrain=150$, corresponding to 150 HF model simulations to train the Gaussian Process model. In all problems we investigated, a choice of~$\ntrain\in\mathbb{N}:[50,200]$ offered a good balance between accuracy and performance. \Cref{fig: ylf_yhf_cylinder} shows the HF and LF model dependency in~$\ssp{\yhf\times\ylf}$ along with the GP-based probabilistic model that would result without~$\gamma_i$. The Gaussian Process model in~$\ssp{\ylf\times\yhf}$, shown in \Cref{fig: ylf_yhf_cylinder}, does not sufficiently explain the complex, non-Gaussian nature of the Monte-Carlo reference, shown by gray dots~(normally not available). Introducing a further dimension~$\gamma_i$ leads to a higher dimensional space where a GP can better explain the data. The transition from~$\ssp{\yhf\times\ylf}$ to~$\ssp{\yhf\times\ylf\times\gamma_1}$ did not require any further HF model evaluation and followed the procedure described in \Cref{sec: informative_features} so that a reduction in the overall approximation error is possible \emph{without further computational efforts}.
 
\begin{figure}[htbp]
 \centering
  \includegraphics[scale=0.58]{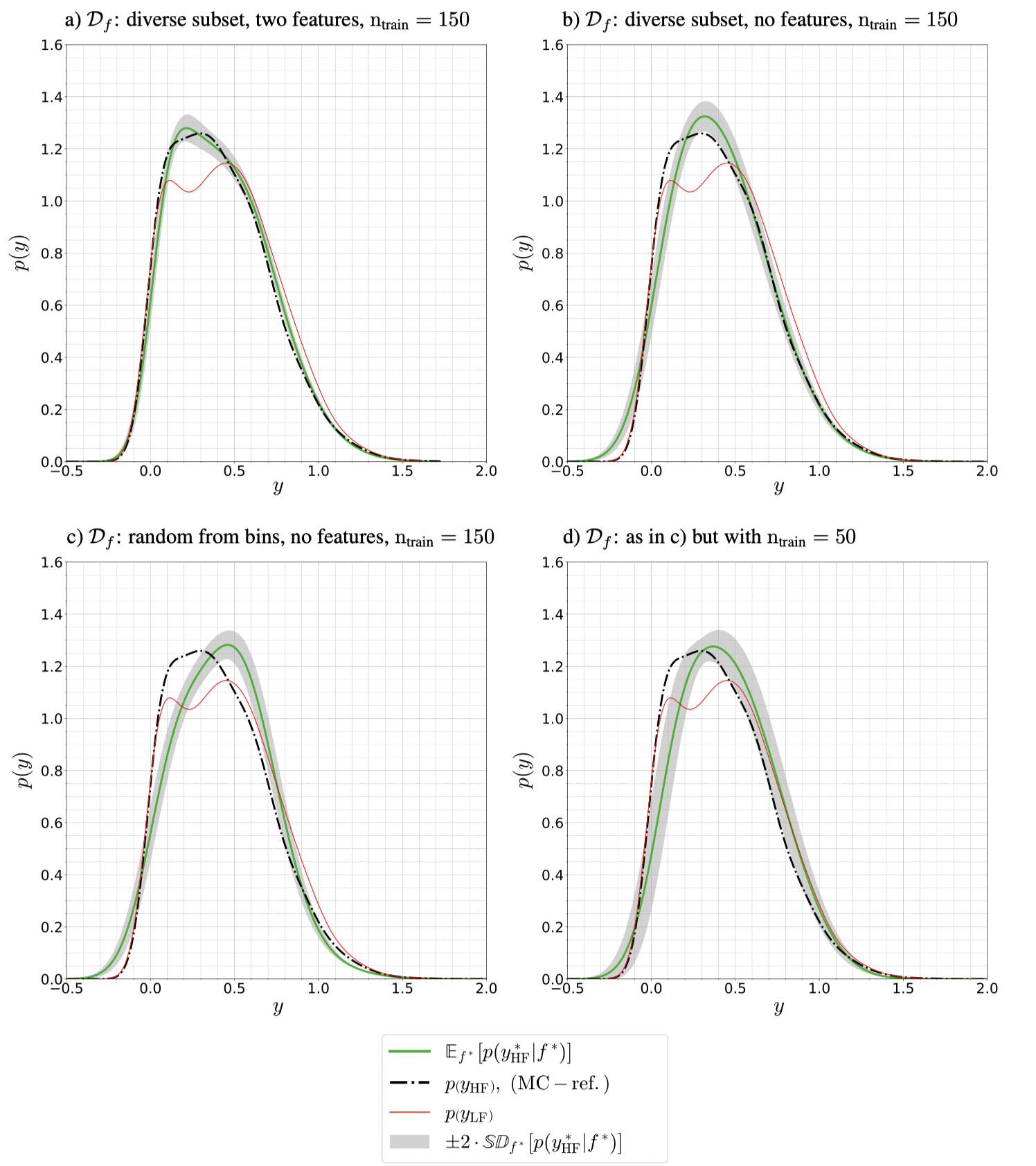}
  \caption{Comparison of the output distributions~$\pdf{y}=\pdf{C_{\text{L,max}}}$ for the maximal lift coefficient in the flow around a cylinder problem. The high-fidelity Monte-Carlo reference~$\pdf{\yhf}$ is shown as a black dash-dot line, the low-fidelity solution~$\pdf{\ylf}$ is given in red color and the BMFMC mean predictions~$\Ex{\ftest}{\pdf{\ytesthf|\ftest,\Ds}}$ is shown in green, along with~$\pm 2$ standard deviation credible intervals of the prediction, shown in gray.} 
  \label{fig: pdfs_cylinder}
 \end{figure}

The resulting approximation for the HF response~$\pdf{\yhf|\Ds}$  is shown in \Cref{fig: pdfs_cylinder}a) to d). 
\Cref{fig: pdfs_cylinder}a) shows the best BMFMC prediction for the HF output density, using~$\ntrain=150$, and two additional input features~$\gamma_1$ and~$\gamma_2$. We determined the informative input features with the heuristic from Section \ref{sec: meaning_features} and specifically Equation \eqref{eqn: LF_sensitivityb}. In this case, the two input features are the two most sensitive input dimensions for the LF model. The training points~$\Ds$ were selected by choosing a diverse subset in~$\ssp{\featExt}$ with~$\mathrm{n}_{\opsExt}=5$. The credible intervals on the densities were computed using Equation~\eqref{eqn: cond_variance}, respectively the square root of Equation~\eqref{eqn: cond_variance} for the standard deviation, and provide an estimate for the uncertainty in the density prediction. The maximum lift coefficient's LF density~(red line) shows a bimodal characteristic that cannot be found in the HF reference density. The BMFMC prediction for~$\ntrain=150$ without informative LF features in \Cref{fig: pdfs_cylinder}b) already resulted in excellent predictions. The addition of two informative features in \Cref{fig: pdfs_cylinder}a) gave even better predictions for the distribution's tails and resulted in slightly lower predictive variance~(narrower credible intervals) and was in almost perfect agreement with the Monte-Carlo density estimate~(dashed black line) that used~$\nsample=10000$ HF evaluations. Figures \ref{fig: pdfs_cylinder}c) and d) used a different strategy to select the training data set~$\Ds$: The outcomes of the LF Monte-Carlo simulation~$\Ytestlf$ were separated into 25 bins, and then an equal amount of training candidates was randomly selected from each bin to define~$\Ds$. Even though this strategy covers~$\ylf$ efficiently, the input~$\bx$ was not sufficiently covered by the training data, resulting in worse predictions with higher predictive variance. \Cref{fig: pdfs_cylinder}d) demonstrates that the predictive variance~(credible intervals on the HF density estimates) provides larger credible intervals for a smaller training data size of~$\ntrain=50$. 

\begin{figure}[htbp]
\centering
  \includegraphics[scale=1.3]{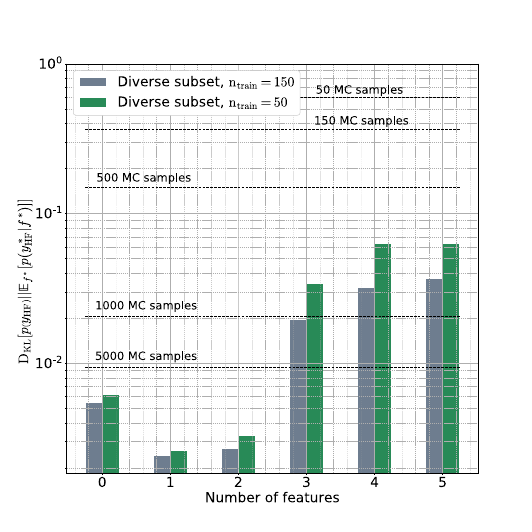}
  \caption{Kullback-Leibler divergence between the Monte-Carlo reference solution~$\pdf{\yhf}$ and the Bayesian prediction~$\Ex{\ftest}{\pdf{\ytesthf|\ftest}}$ for different number of features~$\bs{\gamma}$ in~$\ops$~(logarithmic scale). The horizontal dashed lines mark the KLD for one Monte-Carlo estimate with fewer points towards the Monte-Carlo estimate with~$\nsample=10000$. The Monte-Carlo density estimates yield a significantly higher KLD than the BMFMC estimates.} 
  \label{fig: reduction_features_cylinder}
\end{figure}

\FloatBarrier
\begin{remark}[Kullback-Leibler divergence as error measure]
To measure the accuracy of the predictive HF distribution~$\Ex{\ftest}{\pdf{\ytesthf|\ftest}}$, we define an absolute error measure in terms of the Kullback-Leibler divergence~(KLD) in Equation~\eqref{eqn: absolute_error} towards the HF Monte-Carlo density estimate~$\pdf{\yhf}$, which was calculated with a Gaussian kernel density estimation with bandwidth optimization, using~$\nsample=10000$:
\begin{align}
    \label{eqn: absolute_error}
    \begin{split}
    \eabs:&=\DKL{\pdf{\yhf}}{\Ex{\ftest}{\pdf{\ytesthf|\ftest}}}\\
    &=\int\limits_{-\infty}^{\infty}\pdf{\yhf}\ln\left(\frac{\pdf{\yhf}}{\Ex{\ftest}{\pdf{\ytesthf|\ftest}}}\right) d \yhf
    \end{split}
\end{align}
The KLD is an asymmetric similarity measure between two probability densities. Two identical distributions would result in a KLD value of zero and a discrepancy between the densities in KLD values greater than zero. \Cref{fig: reduction_features_cylinder} shows the performance of the Bayesian multi-fidelity approach using the KLD over an increasing number of features~$\gamma_i$ and two different training sizes~$\ntrain$. To give a reference for the KLDs of the BMFMC solution, we provide the KLD for Monte-Carlo-based density estimates using a lower number of sample points~(horizontal dashed lines) compared to the Monte-Carlo reference using~$\nsample=10000$.
\end{remark}

\FloatBarrier

The KLD for the BMFMC predictions with zero to two additional LF features, which only required 50 HF model simulations, lies far below the KLD that was reached with the density estimate using 5000 HF model evaluations. We also plotted the reference for a density estimate using only 50 HF simulations, resulting in a considerably worse prediction than BMFMC~(please note the logarithmic scale). The best BMFMC predictions were made with only one additional informative input feature~$\gamma_1$. Using one or two additional input features significantly reduces the KLD, without any additional computational cost. For the small training data set~$\Ds$ with~$\ntrain=50$ and~$\ntrain=150$, the introduction of more than two additional features led to a substantial increase in the approximation error due to the curse of dimensionality. The diverse subset strategy for training point selection resulted in considerably better estimates than a random training point selection (not shown).

In conclusion, a computational cost comparison of \emph{BMFMC} with a standard Monte-Carlo procedure, as presented in Equation~\eqref{eqn: speed-up}, 
resulted in an overall speed-up factor of roughly 7.1,
using~$\mathrm{N}_{\text{MC}}=10000$,~$\nsample=150$ and~$f_{\text{HF/LF}}=8$, for the problem at hand. 
In the case of~$\nsample=50$, the speed-up factor was roughly 7.7. The deployed LF model, which was created by changing the degree of the polynomial Ansatz function, is only to be understood as a proof of concept. Much higher speed-ups are possible when further strategies of \Cref{sec: aspects_implementation} are combined, or even a simplified physical model is applied.

\subsection{Stochastic Fluid-Structure Interaction -- Bending Wall in a Channel Flow}
\label{sec:stochastic_fsi}
\begin{figure}[htbp]
 \centering
  \includegraphics[scale=1.1]{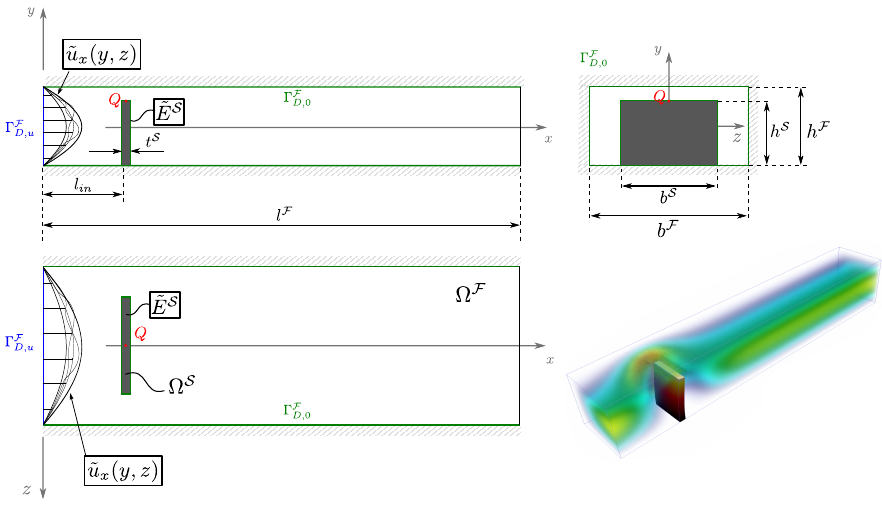}
  \caption{Fluid-structure interaction problem of an elastic wall in a channel flow~\cite{gerstenberger2010embedded} subject to a random inflow boundary condition and uncertain wall elasticity. Random fields, respectively variables are shown in boxes and have an additional tilde superscript. A no-slip boundary condition is present at~$\Gamma_{D,0}^{\mathcal{F}}$. The quantity of interest is the wall deflection in the x-direction at point~$Q$~(shown in red). The Dirichlet boundaries for the fluid and the structure domain are shown in blue, respectively green. The subfigure on the bottom right shows an exemplary three-dimensional result of the flow and displacement field in a graphical projection.} 
  \label{fig: flow_channel}
 \end{figure}

In the second numerical example, we are interested in uncertainty propagation for a 3D fluid-structure interaction~(FSI) problem of a bending wall in a channel flow. The example is motivated by our earlier work on FSI solvers~\cite{gerstenberger2010embedded} and is depicted in \Cref{fig: flow_channel}. The fluid domain~${\Omega}^\mathcal{F}$ is given by a flow-channel with rectangular cross-section of width~$b^{\mathcal{F}}=1.0$, height~$h^{\mathcal{F}}=0.5$ and length~$l^{\mathcal{F}}=3.0$, while the structure domain~$\Omega^{\mathcal{S}}$ is represented by an elastic wall of thickness~$t^{\mathcal{S}}=0.05$, width~$b^{\mathcal{S}}=0.6$ and height~$h^{\mathcal{S}}=0.4$. We assume the flexible wall to be clamped to the channel floor at~$y=-\frac{h^{\mathcal{F}}}{2}$. The distance between the wall's centerline and the left boundary of the fluid domain is~$l_{\text{in}}=0.5$.
The fields are modeled using a hyper-elastic neo-Hookean constitutive law for the structural domain and incompressible Newtonian fluid in the fluid domain. We solve the fluid-structure interaction problem efficiently with a monolithic coupling scheme implemented in our in-house multi-physics finite element code \emph{BACI}~\cite{wall2017baci}. The interested reader is referred to~\cite{verdugo2016unified} for further details on n-field monolithic solvers. In our configuration, the fluid field was chosen as the master field in the dual mortar formulation for the interface coupling~\cite{kloppel2011fluid}. The Dirichlet boundary conditions are formulated on~$\Gamma_D^{\mathcal{F}}=\Gamma_{D,u}^{\mathcal{F}}\cup\Gamma_{D,0}^{\mathcal{F}}$ as follows:
\begin{align}
\label{eqn: Dirichlet_BC2}
\begin{split}
\Gamma_{D,u}^{\mathcal{F}}&: \quad \mathbf{u}=
\begin{bmatrix}
u_x(y,z)\\
0\\
0
\end{bmatrix}\quad \text{inflow B.C.}\\
\Gamma_{D,0}^{\mathcal{F}}&: \quad \mathbf{u}=
\begin{bmatrix}
0\\
0\\
0
\end{bmatrix}\quad \text{no slip B.C.}\\
\end{split}
\end{align}
An overview of detailed material and fluid properties is provided in Appendix \ref{sec: details_bending_wall} in \Cref{tab: simulation_properties}.

The uncertainty propagation problem is stated as follows: The model is subject to two sources of input uncertainties: A random inflow boundary condition~$\tilde{u}_x(y,z)$~(random field with high stochastic dimension) and an uncertain Young's modulus~$\tilde{E}^{\mathcal{S}}$~(random variable) for the elastic wall. We are interested in the distribution of the x-displacement of  point~$Q$ on top of the elastic wall. To be compliant with our notation in previous sections, we summarize the distribution of the inputs by~$\pdf{\bx}$ and the response distribution for the QoI of the high-fidelity computer model is denoted by~$\pdf{\yhf}$. 

\begin{figure}[htbp]
 \centering
  \includegraphics[scale=0.3]{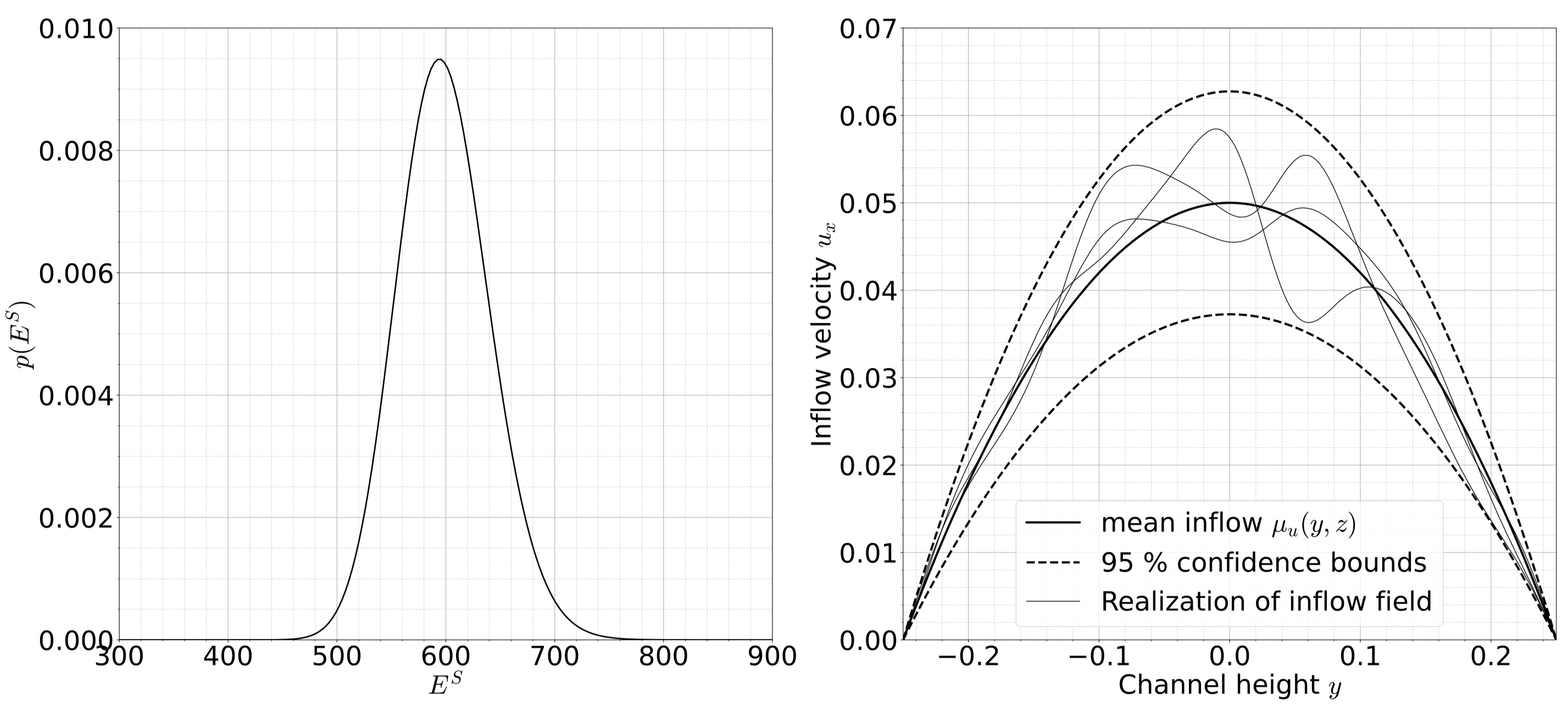}
  \caption{Log-normal density function of the Young's modulus~(left) and visualization of a 2D cross-section at~$z=0$ of the 3D-random inflow field~(right).} 
  \label{fig: distributions}
 \end{figure}

The \textbf{Young's modulus}~$E^{\mathcal{S}}$ of the elastic wall is modeled as a random variable with a log-normal distribution~$\pdf{E^{\mathcal{S}}}~=~\mathcal{LN}_{E^{\mathcal{S}}}(\mu_E,\sigma_E^2)$ to constrain realizations to~$\real{+}$. The distribution parameters~$\mu_E$ and~$\sigma_E^2$ are chosen so that the Young's modulus's mean value is 600 with a standard deviation of 7\% of its mean. The \textbf{random inflow}~$u_x(y,z)$ is realized as a non-stationary Gaussian random field with parabolic mean function on~$\Gamma_{D,u}^{\mathcal{F}}$, analogously to the previous numerical example. Again we can factorize the process to:
\begin{align}
\label{eqn: random_inflow}
u_x(y,z)\sim \mathcal{GP}\left(\mu_u(y,z),\tilde{k}(y,{y}')\right)=\mu_u(y,z)+\mathcal{GP}\left(0,\tilde{k}(y,{y}')\right)
\end{align}
The parabolic mean function was taken from the deterministic problem in~\cite{gerstenberger2010embedded} and is given in \Cref{tab: simulation_properties} along with further properties of the stochastic problem. Densities of the Young's modulus and the resulting random field for the uncertain inflow are visualized in \Cref{fig: distributions}. In the numerical implementation, the realizations of the random inflow were discretized at 200 points so that the stochastic dimension of the problem was~$\dim(\bx)=201$. 

For the stochastic FSI problem, we investigated three different low-fidelity model versions besides the HF model regarding their impact on the overall prediction quality. The problem was solved with the continuous finite element method: The high-fidelity model used for the fluid-domain~$\Omega^{\mathcal{F}}$ 22704 equal-order Hex-8 finite elements with residual-based stabilization. The structure domain~$\Omega^{\mathcal{S}}$ of the HF model used 1536 HEX-8 F-Bar finite elements~\cite{gerstenberger2010embedded,de1996design}. The three low-fidelity model variants were constructed using pure numerical relaxation as described in \Cref{sec: aspects_implementation}:
\begin{description}
\item[LF 1] The first LF model~(LF 1) was designed with 100 times looser solver tolerances for the fluid-structure coupling and a two times larger time step size resulting in an overall speed-up factor of four.
\item[LF 2] The second LF model~(LF 2) was constructed by spatial coarsening to 2838 fluid elements and 192 structural elements while leaving other numerical settings untouched, leading to a speed-up factor of ten. 
\item[LF 3] Finally, the third LF model~(LF 3) combined the relaxations of LF 1 and LF 2 and led to a speed-up factor of roughly 28.
\end{description}

The procedure for the Bayesian multi-fidelity uncertainty propagation scheme was then conducted as in the previous example. We first created an input sample matrix~$\Xtest\sim\pdf{\bx}$ with~$\nsample=7000$ samples using random number generators for the log-normal distribution and a Cholesky decomposition from Equation~\eqref{eqn: realizations_inflow} to generate sample functions for the random inflow boundary condition. Compared to the previous demonstration, a smaller sample size was chosen, as the smaller variance of the response distribution converged faster. Afterwards, we ran the realizations for~$\bx$ on all LF models to get the three response vectors~$\Ytestlf$. Features and inputs~$\Xtrain$ were chosen in the same procedure as the previous numerical demonstration. The difference is that the number of training points was further reduced to~$\ntrain=50$ HF simulations to demonstrate the capabilities of the multi-fidelity approach even for a very small number of high-fidelity simulations. In case the predictive statistics for the HF output density show too high variance according to Equation~\eqref{eqn: cond_variance}, more training points can be calculated. From a practical perspective, we suggest constructing subsets of space-filling training designs~$\Xtrain_1\subset\Xtrain_2\subset\dots\Xtrain_m$ following the procedure in \Cref{sec: informative_features}. Afterwards, we can start with the smallest subset~($\Xtrain_1$) and choose larger sets, reusing the previous simulations, in case the predictive variance of the HF density is still too high.

\begin{figure}[htbp]
 \centering
  \includegraphics[scale=0.22]{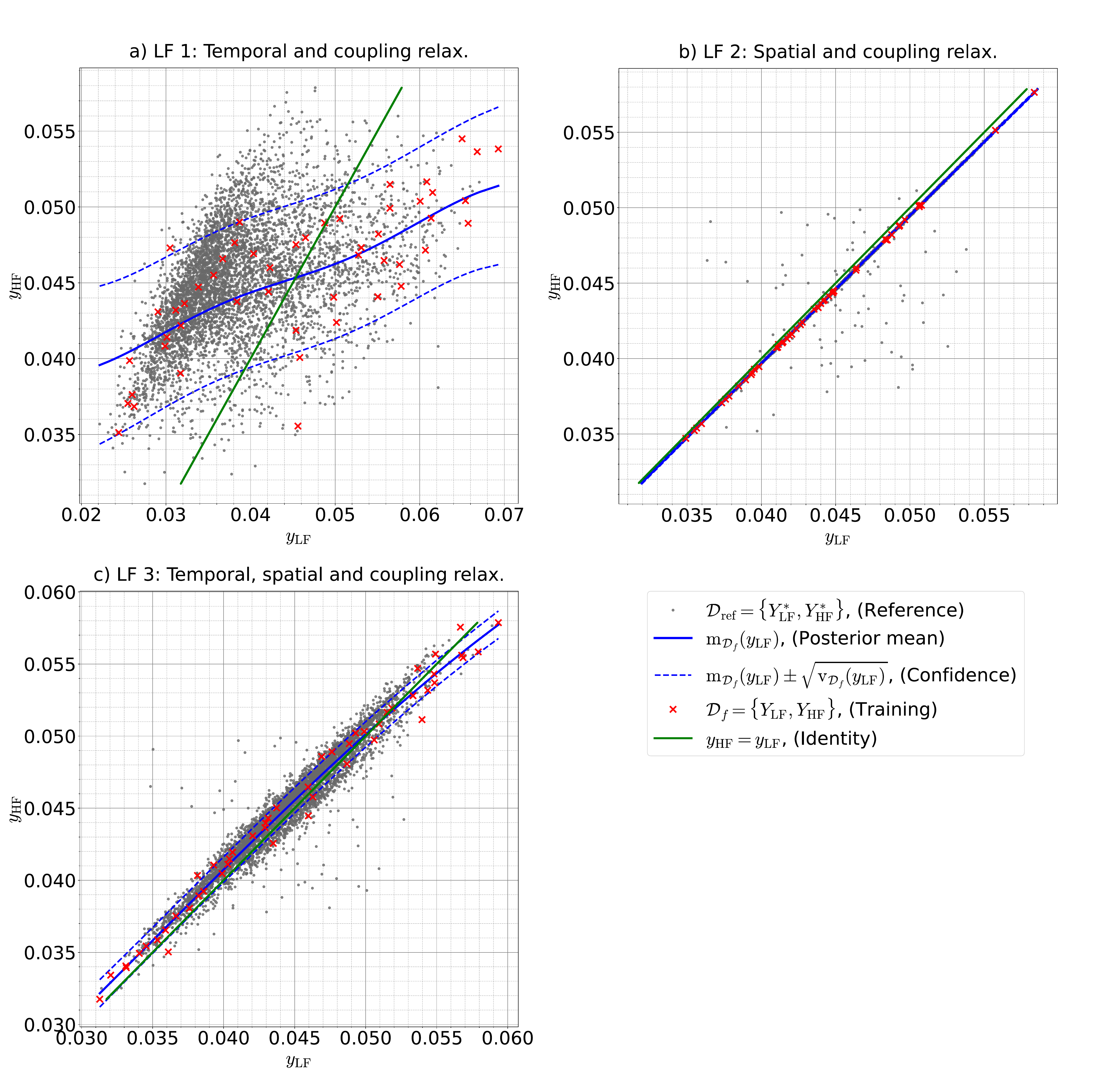}
  \caption{Stochastic fluid-structure interaction problem: HF and LF model outputs for the wall deflection, demonstrated for the three low-fidelity models LF 1, LF 2, and LF 3. The data of the Monte-Carlo reference~(usually not available) is shown as gray dots. Red crosses show training points~($\ntrain=50$) for the Gaussian Process. The blue lines indicate the posterior mean function~$\mf{\ylf}$ and~$\pm 1-\sigma$ standard deviation of the trained Gaussian Process.} 
  \label{fig: fsi_ylf_yhf}
 \end{figure}

\Cref{fig: fsi_ylf_yhf}a) shows the stochastic dependency between the HF model and LF 1, created by relaxation of the time discretization and coupling settings. Please see also Appendix \ref{sec:average_expectation} and specifically Figure \ref{fig:latent_feature} for an illustration of the extended multi-fidelity space, using one informative feature $\gamma_1$ along with the output of LF 1.
The conditional dependency has a non-Gaussian noise structure and strong nonlinearities. The pure spatial relaxation in LF 2 is displayed in \Cref{fig: fsi_ylf_yhf} b). The data points are very close to the identity of~$\yhf=\ylf$~(shown in green), along with some scattered points for which the LF simulation deviated from the HF simulation. The combination of relaxations from LF 1 in Figure \ref{fig: fsi_ylf_yhf} a) and LF 2 in Figure \ref{fig: fsi_ylf_yhf} b) is shown as LF 3 in Figure \ref{fig: fsi_ylf_yhf} c). Despite having the highest numerical relaxation and, therefore, the highest computational speed-up, the LF 3 model results in a less noisy model dependency structure in~$\ssp{\yhf\times\ylf}$ when compared to the LF 1 model. A possible explanation for this effect might be the coarser mesh's smoothing property, which dampens the model discrepancy over the input space. However, a detailed investigation of the effect is outside the scope of this paper.

\begin{figure}[htbp]
 \centering
  \includegraphics[scale=0.58]{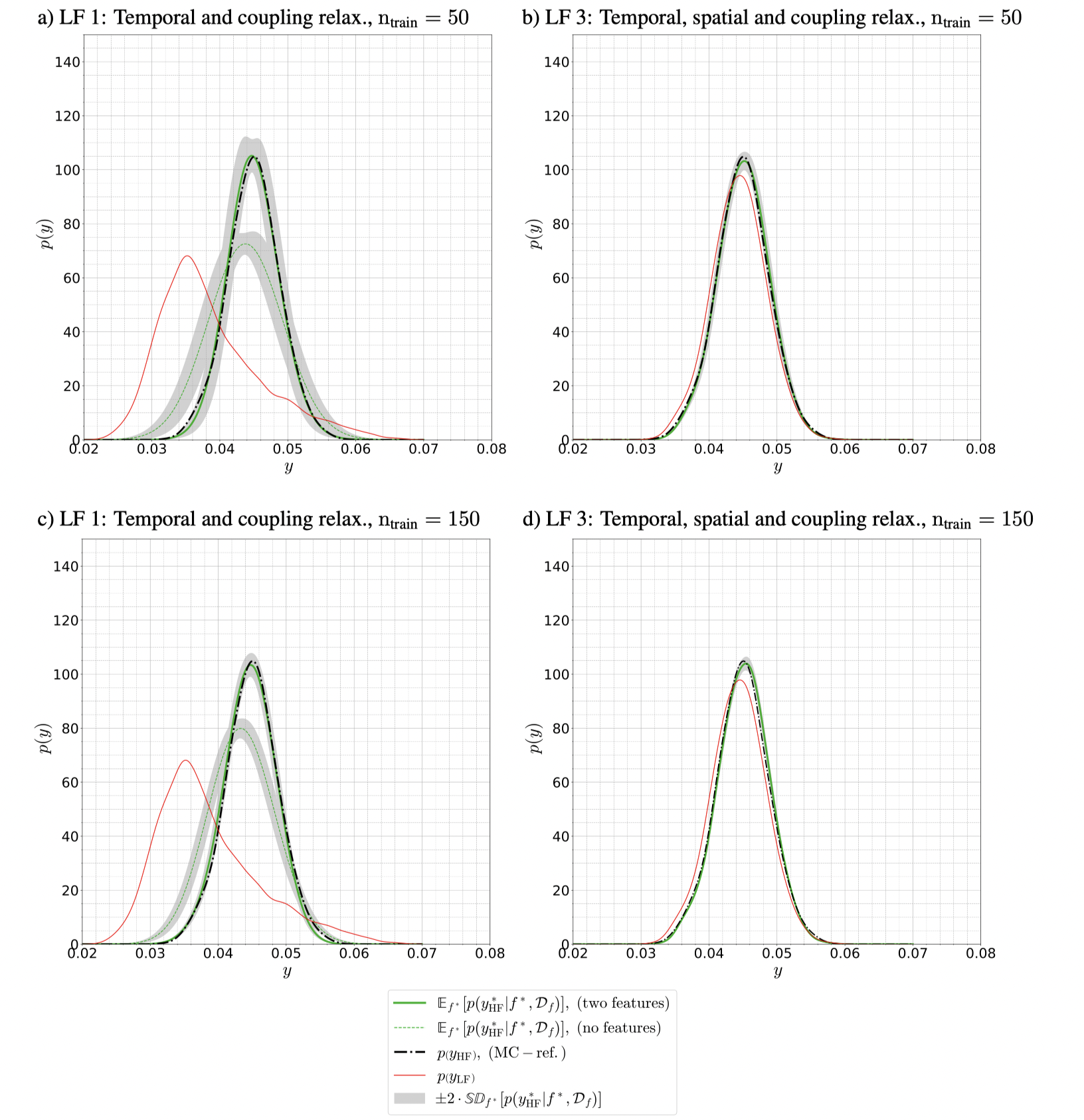}
  \caption{Comparison of the output densities~$\pdf{y}$ for the bending wall in a channel flow problem. The quantity of interest is the x-deflection point Q. The high-fidelity Monte-Carlo reference density~$\pdf{\yhf}$ is shown as a black dashed line, the low-fidelity density~$\pdf{\ylf}$ is given in red, and the Bayesian predictions~$\Ex{\ftest}{\pdf{\ytesthf|\ftest,\Ds}}$, in green along with their credible intervals, shown in gray. Figure a) shows the output densities when LF 1 was deployed in the multi-fidelity approach and Figure b) when LF 3 was deployed each for~$\ntrain=50$. Figure c) and d) show the predictions for~$\ntrain=150$.}
  \label{fig: pdfs_fsi}
\end{figure}

\Cref{fig: pdfs_fsi} presents the resulting density predictions for the HF output along with the MC reference~$\pdf{\yhf}_{\text{MC}}$ and the output densities for LF 1 in \Cref{fig: pdfs_fsi} a) and LF 3 in \Cref{fig: pdfs_fsi} b), respectively. 
Even though the LF 1 model led to a very noisy and non-Gaussian model dependency with the HF model, as shown in \Cref{fig: fsi_ylf_yhf} a), the multi-fidelity approach BMFMC was able to predict the HF output density nearly perfectly with as little as 50 HF simulation runs. We emphasize that \BMFMC achieved a very accurate prediction, even though the LF model provided a very inaccurate output density (shown in red). Additionally, we present the prediction quality of BMFMC without the use of additional informative input features~(green dashed line), which results in considerably worse predictions due to a high modeling error when assuming a Gaussian noise structure between HF and LF 1 in~$\ssp{\ylf\times\yhf}$. In fact, without the use of~$\bs{\gamma}$ the modeling error of the GP is so high that the HF reference density did not lie within the predicted credible intervals, even for an increase in training data as demonstrated in \Cref{fig: pdfs_fsi}c). The introduction of only one additional feature~$\gamma_1$ removed this problem, and the reference solution is always within the credible intervals. \Cref{fig: pdfs_fsi} shows the superior result for two additional features~$\gamma_1$ and~$\gamma_2$ that also gave tighter credible intervals than the prediction with only~$\gamma_1$.
In the case of the LF 3 model, which combined spatial and temporal relaxation, BMFMC predicted the reference density nearly perfectly with or without additional features.

Finally, we investigate the computational speed-ups reached with \emph{BMFMC} in comparison to a standard Monte-Carlo strategy. We use the presented speed-up definition in Equation~\eqref{eqn: speedup} and~\eqref{eqn: speed-up} to calculate the speed-up factor. In our simulations we set~$N_{\text{HF}}=N_{\text{LF}}$ and the resulting speed-up factors are summarized in \Cref{tab: speed-up}.
 \begin{table}[htbp]
  \caption{Comparison of computational speed-ups for the generalized \emph{BMFMC} approach with a standard Monte-Carlo approach for uncertainty quantification.}
  \label{tab: speed-up}
  \centering
  \def\arraystretch{1.5}
  \begin{tabular}{lllll}
    \toprule
    LF model&$f_{\text{HF/LF}}$&$\mathrm{N}_{\text{MC}}$&$\ntrain$&speed-up \BMFMC\\
    \midrule
    LF 1&4.5&7000&50&4.4\\
    LF 2&10&7000&50&9.3\\
    LF 3&28&7000&50&23.3\\
    \bottomrule
  \end{tabular}
\end{table}
Only based on pure numerical relaxation, \emph{BMFMC} performed roughly 23 times faster than the Monte-Carlo benchmark while reaching comparable accuracy. We want to emphasize that we did not even fully exhaust the potential in the numerical relaxation, i.e., the floating-point precision in the simulation was kept untouched. Furthermore, a vast speed-up potential is still available in terms of simplified physics for the model so that the discussed problems do by far not represent the full potential of \emph{BMFMC}.

\section{Conclusion}
\label{sec:conclusion}
In this contribution we presented a generalized version of Bayesian multi-fidelity Monte-Carlo~(\emph{BMFMC}) for uncertainty quantification. Given a high-fidelity model~(HF), a low-fidelity model~(LF), and input uncertainties of the parametric input uncertainties~$\bx$ with a density~$\pdf{x}$, \BMFMC yields a very accurate approximation for the sought high-fidelity output density~$\pdf{\yhf}$ and requires only sampling on the LF model at a reduced computational cost. Only a small amount of HF evaluations is necessary to learn a nonlinear approximation of the \emph{multi-fidelity conditional distribution}~$\pdf{\yhf|\ylf}$. The curse of dimensionality is circumvented by filtering the high dimensional input $\bx$ through the LF model~$\ylf(\bx)$, such that no~(probabilistic) surrogate concerning the inputs needs to be constructed. This allows us to address UQ problems in high dimensions without losing accuracy in the formulation.

The continuous Bayesian multi-fidelity formulation in Equation \eqref{eqn: mf_uq}, before any discretization and approximation steps, is mathematically exact and recovers the actual high-fidelity output density $\pdf{\yhf}$. We have identified two error sources in the discretized, practically relevant version of our approach, stemming from a) the model class for the multi-fidelity conditional~$\pdf{\yhf|\ylf}$~(\textbf{error 1}) and b) the finite~(and generally small) amount of training data~(\textbf{error 2}). We demonstrated how the two error sources can be controlled to enable accurate uncertainty quantification in high stochastic dimensions using very few high-fidelity simulation runs. In particular, we note that the overall error in the proposed \emph{BMFMC} approach is usually much smaller than the approximation error that arises from directly performing UQ on the HF model for the same limited computational budget.

We specifically proposed a strategy to minimize the combination of these errors without increasing the number of HF model runs. To this end, we introduce informative features~$\gamma_i(\bx)$ that complement the LF output~$\ylf$ and lead to an extended space in which the multi-fidelity conditional can be learned with lower error. We show that an optimum of considered dimensions exists in the \emph{small data regime} in which the multi-fidelity relationship should be represented for a minimum error. The framework is furthermore capable of quantifying the uncertainty in the density prediction, which is an approximation for the magnitude of \textbf{error 2} and a further strength of the method.

We presented the specific numerical implementation that uses Gaussian Processes in this contribution. In contrast to popular alternatives such as MLMC methods, we demonstrate that \BMFMC can capture and use nonlinear dependencies between low and high-fidelity models. The generalized \emph{BMFMC} formulation contains other state-of-the-art methods for uncertainty quantification as special cases. We demonstrate that the generalized framework has a drastically increased accuracy and comes at no extra cost. This is especially advantageous for computationally expensive UQ problems with a high stochastic dimension. The Bayesian approach provides credible intervals for the HF density estimate itself so that it is possible to assess the uncertainty of the predictions.

We aimed at the applicability of \emph{BMFMC} towards complex physical models with actual engineering relevance and demonstrated the capabilities and generality of the method on two challenging stochastic fluid-flow and fluid-structure interaction problems with high stochastic input dimensions. We compared the performance of \emph{BMFMC} with Monte-Carlo and demonstrated a speed-up factor of over 23 for a conservatively chosen LF model in combination with \emph{BMFMC}. The speed-ups were achieved by only using simple numerical relaxation of the original problem. We want to emphasize that higher performance gains are possible by, e.g., using simplified physics in the LF model, a simpler geometrical model, or combinations of the numerical relaxations, which was not the focus of this paper.  

Conclusively, we summarize current limitations and practical aspects of \emph{BMFMC}. We first want to consider the (rather theoretic) case that the LF model is independent of the HF model (see \textbf{extreme 1} in Section \ref{sec: multi_fidelity_overview}). The multi-fidelity conditional disorientates then to $\pdf{\yhf|\ops}=\pdf{\yhf|\boldsymbol{\gamma}}$ (if no informative input features are used the expression disorientates to $\pdf{\yhf|\ylf}=\pdf{\yhf}$). The continuous \emph{BMFMC} formulation is still mathematically correct. However, the LF model does not provide any efficiency gains, and the UQ task is shifted to estimating $\pdf{\yhf|\boldsymbol{\gamma}}$, respectively $\pdf{\yhf}$ directly, instead of profiting from a functional dependency. The probabilistic regression model's mean function is a constant function due to the independence of $\yhf$ and $\ylf$. 
Given a limited computational budget and hence a small number of training points (e.g., 50 to 300 HF model runs), a probabilistic regression model can not be expected to accurately infer a potentially complex conditional density in the $\yhf-\ylf$. We emphasize that the same accuracy issues hold for a direct estimation of the HF output density, however, a too simplistic probabilistic regression model, as the employed GP, might have a larger model \textbf{error 1} than a direct kernel density estimation for the HF output density.

In the small data regime, such comparisons are difficult due to a large amount of epistemic uncertainty. More flexible discriminative models might only slightly reduce the model (\textbf{error 1}) but would require more attention regarding regularization. We did not study this influence in the present work but consider it of interest for subsequent research on \emph{BMFMC}.
For now, we can follow the procedure laid out in Section \ref{sec: meaning_features} and \ref{sec: optimal_num_features} and add informative input features $\boldsymbol{\gamma}$ to the multi-fidelity dependency at no computational extra costs.
As informative input features are derived from the LF/HF model input, they share a dependency with the latter and do not only partly compensate \textbf{error 1} but also lead to accuracy gains in \emph{BMFMC}, even if the LF model itself cannot be exploited.

Other practical challenges of the method might arise from the training or optimization of the employed discriminative probabilistic models. In our examples and under the use of GPs we, however, never encountered any problems in this regard. The more challenging aspect in our experience was usually the design of efficient LF models, if not already available. Besides the obvious cases of relaxed numerical discretizations, tolerances and geometries, especially the use of a simplified physical description is very promising for more applied use cases of \emph{BMFMC}. Usually, such simplifications pay off in terms of overall computational speed-up.

In further future work, the presented method will be extended to the entire solution field of an HF model, using more flexible probabilistic modeling approaches. We also think that an extended theoretical investigation on better informative input features can lead to additional improvements in the method. Furthermore, an in-depth investigation of the Bayesian multi-fidelity approach for backward uncertainty propagation~(inverse problems) should yield significant computational speed-ups. Other applications of the method are multi-scale problems where the different physical scales can be interpreted as fidelity levels.

\section*{Acknowledgments}
The research presented in this paper was partly funded by the German Research Foundation~(DFG) under the priority program SPP 1886 "Polymorphic uncertainty modeling for the numerical design of structures" with additional contributions that were funded by the DFG priority program "Software for Exascale Computing"~(SPPEXA). The base version of the software \emph{QUEENS} was provided by courtesy of AdCo Engineering$^{\text{GW}}$ GmbH, which is gratefully acknowledged. We finally want to thank Peter M\"unch for his support on the side of the DG-Navier-Stokes solver, Sebastian Brandst\"ater for his helpful contributions in the software framework \emph{QUEENS} and Gil Robalo Rei for his support in model preparation.

\bibliographystyle{unsrtnat}
\bibliography{mybibfile} 

\appendix
\section{Appendix: Numerical Approximation of Posterior Statistics using Gaussian Processes}
\label{appendix: gp_approx}
In all the analyzed examples, we started with a prior Gaussian Process of the form:
\begin{subequations}
\begin{align}
    \label{eqn: prior_GP}
    \pdf{\fmod}&=\GP{\fmod}{\mathrm{m}\left(\ops \right),\mathrm{k}\left(\ops,{\ops}'\right)},\\
    \mathrm{m}\left(\ops \right)&=\ylf\\
    \mathrm{k}\left(\ops,{\ops}'\right)&={\sigma_0}^2\cdot\exp\left[-\frac{|\ops-{\ops}'{|}^{2}}{2{\ls}^{2}}\right],\label{eqn: s_exp_kernel}
\end{align}
\end{subequations}
where~$\mathrm{m}\left(\ops \right)$ is the prior mean function and~$\mathrm{k}\left(\ops,{\ops}'\right)$ is the prior covariance function which we choose to be the squared exponential covariance function with~$\ls$ being the characteristic length scale and~${\sigma_0}^2$ the signal variance. The prior mean function of the process is set equal to~$\ylf$ as it is assumed that the LF output reflects~$\yhf$ on average. We do not prescribe a functional prior in the direction of~$\gamma_i(\bx)$, such that the prior defaults here to zero. Following our notation, capital letters, e.g.,~${\Gamma_i}^*$, denote the vector of realizations of a random variable, e.g.,~$\gamma_i$, and bold capital letters denote matrices, such a matrix holding column-wise realizations of vector-valued random variables, e.g.~$\Opstrain$, or the covariance matrix~$\Kmat$. The asterisk superscript indicates the large data set derived from the LF Monte-Carlo simulation.
Furthermore, we model a Gaussian likelihood of the data with noise level~$\nvarraw$:
\begin{equation}
    \pdf{\Ytrainhf|\Ftrain,\Opstrain}=\nd{\yhf}{\Ftrain,\nvarraw I},
    \label{eqn: likelihood}
\end{equation}
with~$\Ftrain$ being a particular realization of the GP, evaluated for the feature matrix~$\Opstrain$, where each column corresponds to a training point. The posterior GP~$\pdf{\fmod|\Ds}$, follows then to~\cite{CarlEdwardRasmussen2005}:
\begin{subequations}
\begin{align}
\label{eqn: posterior_GP}
    \pdf{\fmod|\Ds}&=\GP{\fmod|\Ds}{\mf{\ops},\kfun{\Ds}{\ops}{\ops}}\\
     \mf{\ops}&=\mathrm{m}\left(\ops\right)+{\mathrm{\mathbf{k}}}^T\left(\ops\right)\left(\Kmat+\nvar I\right)^{-1}\left(\Ytrainhf-\mathrm{m}(\ops)\right)\\
     \begin{split}
    \kfun{\Ds}{\ops}{\ops}&=\mathrm{k}\left(\ops,{\ops}'\right)\\
    &-\mathrm{k}\left(\ops,\Opstrain\right)\left[\mathrm{k\left(\Opstrain,\Opstrain\right)+\nvar I}\right]^{-1}\mathrm{k}\left(\Opstrain,{\ops}'\right)
   \end{split}
\end{align}
\end{subequations}
Here,~$\Kmat=\mathrm{k}\left(\Opstrain,\Opstrain\right)$ and~$\mathrm{\mathbf{k}}=\mathrm{k}\left(\ops,\Opstrain\right)$ are used for compact notation. Point estimates of the  hyper-parameters of the model~$\param=\{\ls\ ,\sv\ , \sigma_n^2 \}$ are determined by maximizing the marginal likelihood~\cite{CarlEdwardRasmussen2005} and are denoted by~$\hat{\param}$ in the sequel. In addition, we denote the posterior variance of the GP~(i.e., the posterior covariance for~$\ops={\ops}'$) with~${\varf{\ops}=\mathrm{k}_{\Ds}\left(\ops,\ops\right)}$. The involved densities are summarized in \Cref{tab: density_models}:
\begin{table}[htbp]
  \caption{Applied models for the densities in equations~\eqref{eqn: hl_BMFMC1} to~\eqref{eqn: hl_BMFMC3}, respectively applied to the equations~\eqref{eqn: cond_expectation_refined} and~\eqref{eqn: cond_variance_refined}.}
  \label{tab: density_models}
  \centering
  \def\arraystretch{1.5}
  \begin{tabular}{lll}
    \toprule
    Density& Applied model&Description\\
    \midrule
~$\pdf{\ftest|\opstest}$&$\nd{\ftest}{\mathrm{m}\left(\opstest\right),\mathrm{v}\left(\opstest\right)}$&Prior GP evaluated at~$\opstest$\\    
~$\pdf{\ftest|\opstest,\Ds}$&$\nd{\ftest}{\mf{\opstest},\varf{\opstest}}$&Posterior GP evaluated at~$\opstest$\\    
~$\pdf{\ytesthf|\ftest,\opstest}$&$\nd{\ytesthf}{\ftest,\nvar}$&Likelihood of HF data\\
~$\pdf{\ops}$&Only samples available&LF distr.\\
    \midrule
~$\pdf{\ytesthf|\opstest,\Ds}$&$\nd{\ytesthf}{\mf{\opstest},\varf{\opstest}+\nvar}$&Multi-fidelity conditional\\
~$=\int\limits_{\ssp{\ftest}}\pdf{\ytesthf|\ftest,\opstest}\pdf{\ftest|\opstest,\Ds} d\ftest$&&\\
~$\pdf{\ytesthf|\ftest}$& See Equation~\eqref{eqn: hl_BMFMC1} &Random process for HF density\\
~$=\int\limits_{\ssp{\opstest}}\pdf{\ytesthf|\ftest,\opstest}\pdf{\opstest}d \opstest$&&\\
~$\pdf{\ytesthf|\Ds}$&See equations~\eqref{eqn: cond_expectation},~\eqref{eqn: cond_expectation_refined}&Mean estimate for HF density\\
~$=\Ex{\ftest}{\pdf{\ytesthf|\ftest}}$&&(in small data regime)\\
    \bottomrule
  \end{tabular}
\end{table}

The detailed derivation of the mean prediction for the HF~(posterior) density~$\pdf{\yhf}$ from Equation~\eqref{eqn: cond_expectation} follows then to:
\begin{align}
\label{eqn: cond_expectation_refined}
\begin{split}
\Ex{\ftest}{\pdf{\ytesthf|\ftest}}&=\int\limits_{\ssp{\opstest}}\int\limits_{\ssp{\ftest}}\nd{\ytesthf}{\ftest,\nvar}\nd{\ftest}{\mf{\opstest},\varf{\opstest}} \\
&\qquad d \ftest \pdf{\opstest}d\opstest\\
&= \int\limits_{\ssp{\opstest}}\nd{\ytesthf}{\mf{\opstest},\varf{\opstest}+\nvar}\pdf{\opstest}d\opstest\\
&\approx \frac{1}{\nsample}\sum\limits_{j=1}^N \nd{\ytesthf}{\mf{\opstest_j},\varf{\opstest_j}+\nvar}
\end{split}
\end{align}

We calculate the variance of the~$\yhf$ density prediction with respect to the GP realizations~$\ftest$ at~$\opstest$ and make use of the arithmetic for Gaussian distributions to find a semi-analytic formulation for the variance expression up to the integration over~$\opstest$ and~${\opstest}'$, respectively. Again, the outer integrals over~$\ssp{\opstest}$, respectively~${\ssp{\opstest}}'$, have to be solved via Monte-Carlo integration due to the non-Gaussian distributions~$\pdf{\opstest}$, respectively~$\pdf{{\opstest}'}$. The subtrahend~$\left(\Ex{\ftest}{\pdf{\ytesthf|\ftest}}\right)^2$ can be reused from the previous computation in Equation~\eqref{eqn: cond_expectation_refined}. For the subsequent derivation we define the vectors~$\bs{\ytesthf}=[\ytesthf,\ytesthf]^T$ and~$\bs{\ftest}=[\ftest,\ftest]^T$ to denote the support of the multivariate normal distributions, which arise from the multiplication of two univariate normal distributions in the expectation expression~$\Ex{\ftest}{\nd{\ytesthf}{\ftest(\opstest),\nvar}\cdot\nd{\ytesthf}{\ftest({\opstest}'),\nvar}}$~(see~\cite{CarlEdwardRasmussen2005, bertsekas2002introduction} for stochastic calculus):
\begin{align}
    \label{eqn: cond_variance_refined}
    \begin{split}
&\Var{\ftest}{\pdf{\ytesthf|\ftest}}=\\
&=\int\limits_{\ssp{\opstest}}\int\limits_{{\ssp{\opstest}}'} \Ex{\ftest}{\nd{\ytesthf}{\ftest(\opstest),\nvar} \qquad\cdot\nd{\ytesthf}{\ftest({\opstest}'),\nvar}}\pdf{\opstest}\pdf{{\opstest}'}d{\opstest}' d\opstest-\left(\Ex{\ftest}{\pdf{\ytesthf|\ftest}}\right)^2\\
&=\int\limits_{\ssp{\opstest}}\int\limits_{{\ssp{\opstest}}'}\int\limits_{\ssp{\ftest}} \nd{\mathbf{\ytesthf}}{\begin{bmatrix}\ftest(\opstest)\\ {\ftest}'({\opstest}')\end{bmatrix},\begin{bmatrix}\nvar &0\\ 0 & \nvar \end{bmatrix}}\\
&\qquad\qquad\cdot \nd{\bs{\ftest}}{\begin{bmatrix}
\mf{\opstest}\\
\mf{{\opstest}'}
\end{bmatrix},\begin{bmatrix}
\varf{\opstest} & \kfun{\Ds}{\opstest}{\opstest} \\
\kfun{\Ds}{\opstest}{\opstest}&\varf{{\opstest}'}
\end{bmatrix}
}d\ftest\\
&\qquad\qquad \cdot\pdf{\opstest}\pdf{{\opstest}'}d{\opstest}' d\opstest-\left(\Ex{\ftest}{\pdf{\ytesthf|\ftest}}\right)^2\\
&=\int\limits_{\ssp{\opstest}}\int\limits_{{\ssp{\opstest}}'}\nd{\mathbf{\ytesthf}}{\begin{bmatrix}
\mf{\opstest}\\
\mf{{\opstest}'}
\end{bmatrix},\begin{bmatrix}
\varf{\opstest}+\nvar & \kfun{\Ds}{\opstest}{\opstest} \\
\kfun{\Ds}{\opstest}{\opstest} & \varf{{\opstest}'}+\nvar
\end{bmatrix}}\\
&\qquad\qquad\cdot \pdf{\opstest}\pdf{{\opstest}'}d{\opstest}' d\opstest -\left(\Ex{\ftest}{\pdf{\ytesthf|\ftest}}\right)^2\\
&\approx\frac{1}{\nsample^2}\sum\limits_{i,j=1}^{\nsample}
\nd{\mathbf{\ytesthf}}{\begin{bmatrix}
\mf{\opstest_i}\\
\mf{\opstest_j}
\end{bmatrix},\begin{bmatrix}
\varf{\opstest_i}+\nvar & \kfun{\Ds}{\opstest_i}{\opstest_j} \\
\kfun{\Ds}{\opstest_j}{\opstest_i} & \varf{\opstest_j}+\nvar
\end{bmatrix}}\\
&\qquad\qquad-\left(\Ex{\ftest}{\pdf{\ytesthf|\ftest}}\right)^2
\end{split}
\end{align}

\section{Appendix: Average Variance of Extended Space}
\label{sec:average_expectation}
It is to be proven, that the average variance of an extended distribution~$\Ex{b,c}{\Var{a|b,c}{\pdf{a|b,c}}}$ is smaller or equal than the average variance of~$\Ex{b}{\Var{a|b}{\pdf{a|b}}}$:
\begin{equation}
  \label{eqn:to_be_proven}
\text{To be proven: }  \Ex{b,c}{\Var{a|b,c}{\pdf{a|b,c}}} \le \Ex{b}{\Var{a|b}{\pdf{a|b}}}
\end{equation}
We use the following two \emph{laws} to proof Equation~\eqref{eqn:to_be_proven}. First, we use the \emph{law of total variance}:
\begin{equation}
  \label{eqn:law_of_total_variance}
  \Var{a|b}{\pdf{a|b}} = \Ex{c|b}{\Var{a|b,c}{\pdf{a|b,c}}} + \Var{c|b}{\Ex{a|b,c}{\pdf{a|b,c}}} 
\end{equation}

Second, we use the \emph{law of iterated expectations}:
\begin{equation}
  \label{eqn:law_of_iterated_expectation}
\Ex{a,b}{\pdf{a,b}}=\Ex{b}{\Ex{a|b}{\pdf{a,b}}}
\end{equation}

Taking the expectation of the Equation~\eqref{eqn:law_of_total_variance} w.r.t.~$\pdf{b}$ results in:
\begin{align}
\label{eqn:proof}
\begin{split}
\Ex{b}{\Var{a|b}{\pdf{a|b}}}&= \Ex{b}{\Ex{c|b}{\Var{a|b,c}{\pdf{a|b,c}}}} + \Ex{b}{\Var{c|b}{\Ex{a|b,c}{\pdf{a|b,c}}}}\\
&=\Ex{b,c}{\Var{a|b,c}{\pdf{a|b,c}}} + \underbrace{\Ex{b}{\Var{c|b}{\Ex{a|b,c}{\pdf{a|b,c}}}}}_{>0}\\
&\Rightarrow \Ex{b,c}{\Var{a|b,c}{\pdf{a|b,c}}} \le \Ex{b}{\Var{a|b}{\pdf{a|b}}}
\end{split}
\end{align}
In Equation~\eqref{eqn:proof} we used the \emph{law of iterated expectations} in the second line. 
Please note, that~$\Var{a|b,c}{\pdf{a|b,c}} \le \Var{a|b}{\pdf{a|b}}, \ \forall a,b$ usually does not hold.

\begin{figure}[htbp]
  \centering
  \includegraphics[scale=0.9]{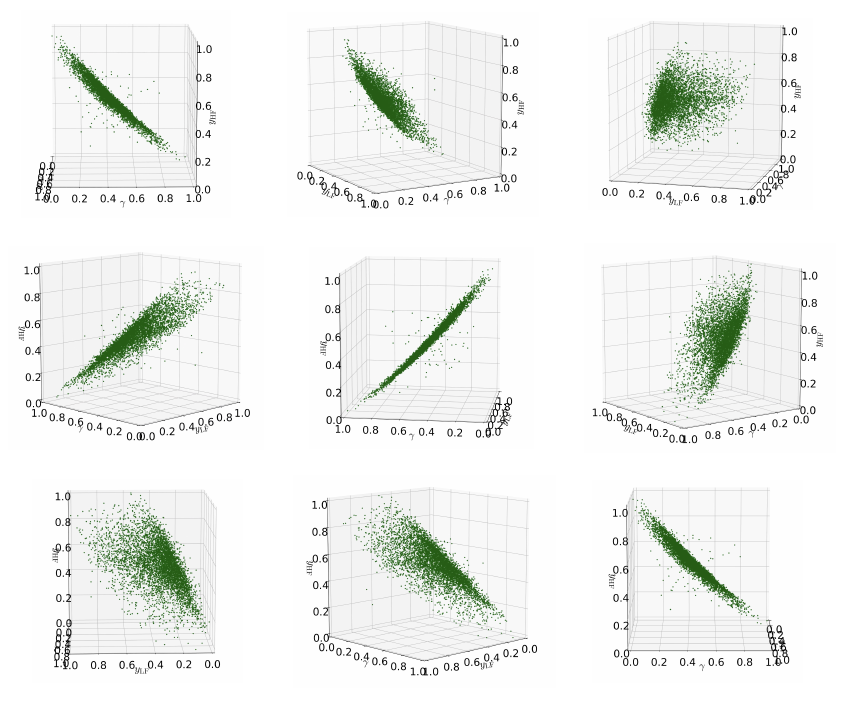}
  \caption{Illustration of the multi-fidelity dependency's \emph{simpler} data structure in the extended space with one additional informative input feature $\gamma_1$. Shown is the normalized multi-fidelity Monte-Carlo reference data using the LF 1 (temporal and coupling relaxation) of the stochastic fluid-structure interaction problem in Section \ref{sec:stochastic_fsi}. See especially Figure \ref{fig: fsi_ylf_yhf}a) for the data representation without the informative input feature, which is very noisy. The informative input feature $\gamma_1$ was, in this case, taken to be the corresponding sample value of the bending wall's uncertain Young's modulus $\tilde{E}^{S}$.}
  \label{fig:latent_feature}
\end{figure}

\section{Appendix: Exemplary multi-fidelity function}
\label{sec:multi_fid_fun}
The exemplary multi-fidelity function is given by the following expressions:
\begin{subequations}
\begin{align}
  \yhf(\bx) & = 0.2\cdot x_1^{1.2} + 0.23\cdot x_2^{0.8}+ 0.5 \cdot\sin(x_1)\\
  \ylf(\bx) & = 2+0.35\cdot (0.3\cdot x_1^{1.5} + 0.22\cdot x_2^{1.1}) + 0.3\cdot \sin(1.3\cdot x_2 + 0.3)
\end{align}
\end{subequations}

\section{Appendix: Dimensionality Reduction}
\label{sec:dim_reduction}
The input~$\bx$ of the computational models is split into uncorrelated dimensions and correlated dimensions:~${\bx=[\bx_{\text{uncorr}},\bx_{\text{corr},i}]^T}$, with~${i\in\mathbb{N}:[1,\mathrm{n}_{\text{corr}}]}$. In this notation~$\bx_{\text{corr},i}^T$ represent the correlated inputs in form of correlated random variables or random fields. A reduced representation~$\xred$ can be achieved by applying unsupervised dimensionality reduction techniques. We used a truncated Karhunen-Lo\`eve expansion~(KLE) for the random fields, on~$\bx_{\text{corr},i}^T$. We want to emphasize that the dimensionality reduction was not used to generate the realization of random fields deployed in the simulations but is a post-processing step only used in the probabilistic learning approach. The starting point is an eigenproblem for the random field's covariance matrix~$\Kmattest$. The latter is defined by the evaluation of its covariance function for the underlying discretization:
\begin{equation}
    \label{eqn: eigenproblem}
    \Kmattest\bs{v}_{j,i}=\lambda_{j,i} \bs{v}_{j,i}
\end{equation}
The eigenvectors~$\bs{v}_j$ define a complete basis, in which we can represent~$\bx_{\text{corr},i}$ as a linear combination of~$\bs{v}_j$, so that the coefficients of the expansion yields:
\begin{align}
    \label{eqn: truncated_KLE}
    \bs{c}_i&\approx \bs{V}_{\text{trunc},i}^T\cdot\bx_{\text{corr},i}-\bs{m}_{i},\ \text{with}\ i\in\mathbb{N}:[1,\mathrm{n}_{\text{corr}}]
\end{align}
Here,~$\bs{m}_{i}$ is the mean vector of the~$i$-th discretized random field and~$\bs{c}_i$ a vector of coefficients for the truncated basis~$\bs{V}_{\text{trunc},i}$. We truncate the series expansion in Equation~\eqref{eqn: truncated_KLE} when 95\% of the explained variance is reached. The explained variance of the discretized field is defined as:
\begin{equation}
    \label{eqn: explained variane}
   \text{explained variance} \coloneqq \sum\limits_{i=j}^{\mathrm{n}_{\text{trunc}}}\lambda_j/ \sum\limits_{j=1}^{d_{\text{corr}}}\lambda_j, \text{  with }\ d_{\text{corr}}=\dim(\bx_{\text{corr},i})
\end{equation}
Afterwards, we propose to use the vector of KLE-coefficients~$\bs{c}_i$ as a low dimensional feature vector for~$\bx_{\text{corr},i}$. Standardization of each dimension is written in form a standardization operator~$\mathcal{S}$. The reduced input vector then follows to:
\begin{equation}
    \label{eqn: reduced_input}
    \xred\coloneqq \mathcal{S} {[\bx_{\text{uncorr}},\bs{c}_i]}^{T}, \text{ with } i\in\mathbb{N}:[1,\mathrm{n}_{\text{trunc},i}]
\end{equation}
Standardization refers to the individual scaling of the dimensions in~$\xred$ so that their underlying input density has zero mean and a standard deviation of one. 
\section{Appendix: Sub-algorithms of BMFMC}
\label{sec: sub_algorithms}

\begin{algorithm}
\caption{{TrainData}$\left(\Xtest,\Ytestlf,\ntrain,\nsample\right)$}\label{alg: trainingdata}
\begin{algorithmic}[1]
    \State~$\Xred\gets\Xtest$ \Comment{Construct reduced input matrix}
    \State~$\Xred$ = \Call{Standardize}{$\Xred$} \Comment{Standardize input matrix}
\item[]  
\item[// ---------------------------- Define~$\featExt$~(supervised) ---------------------------- //]
\State~$\mathbf{r}=\big\vert\Xred^T\cdot Y_{\text{LF}}^*\big\vert$ \Comment{Calculate corr. coefs.~(Eq.~\eqref{eqn: LF_sensitivityb})}
\For{$i$ to~$\mathrm{n}_{\featExt}$} \Comment{$\mathrm{n}_{\featExt}\in\mathbb{N}:[3,6]$ is a good heuristics}
    \State~$\text{idx}$ = \Call{ReturnIndexMax}{$\mathbf{r}$}\Comment{Dim. in~$\xred$ with max. correl. to~$\yhf$}
    \State~$\Gamma_i^*$ = \Call{SelectColumn}{$\text{idx},\Xred$}\Comment{Select corresp. column in~$\Xred$}
    \State~$\mathbf{r}$ = \Call{SetMaxZero}{$\mathbf{r}$}
\EndFor
\State~$\FeattestExt\gets[\Gamma_i^*], \ i \in\mathbb{N}:[1,\mathrm{n}_{\featExt}]$ \Comment{Construct extended feature space}

\item[]  
\item[// -------------------------------- Select~$\Xtrain$ and~$\Ytrainhf$ --------------------------------- //]
\State~$\FeattrainExt$ = \Call{SelectDiverseSubset}{$\FeattestExt,\ntrain$}
\State~$\Xtrain$ = \Call{GetCorrespondingInput}{$\FeattrainExt$}
\State~$\Ytrainhf=\yhf(\Xtrain)$\Comment{Run HF model for training inp.~$\Xtrain$}
\item[]  
\item[// -------------------------------- Select~$\ops$ and~$\Ds$ --------------------------------- //]
\State~$\ops\gets[\bs{\ylf},\gamma_{i}]$ \Comment{with~$i\in\mathbb{N}:[1,\mathrm{n}_{\bs{\gamma}}]$}
\State~$\Opstrain\gets\ops$
\State~$\Ds\gets[\Opstrain,\Ytrainhf]$\Comment{$\ntrain\ll\nsample$}
\item[]
\State \Return~$\Ds$
\end{algorithmic}
\end{algorithm}

\begin{algorithm}
\caption{{PosteriorStatistics}$\left(\mathcal{GP}_{\fmod|\Ds},\Opstest,\nsample \right)$}\label{alg: posteriorstatistics}
\begin{algorithmic}[1]
    \item[// ---------- Calculate~$\Ex{\ftest}{\pdf{\yhf|\ftest}}$, Equation~\eqref{eqn: cond_expectation} ---------- //] 
    \State~$\bs{m}^*,\bs{v}^*$=\Call{EvaluateGP}{$\mathcal{GP}_{\fmod|\Ds},\Opstest$}
    \State~$\bs{P}_{\yhf}^*$=\Call{Normal}{$\bs{y}_{\text{HF,support}},\bs{m}^*,\bs{v}^*$}\Comment{Column-wise Gauss. samples}
    \State~$\bs{p}_{\yhf,{\mathbb{E}}^{*}}$=$1/\nsample\cdot$\Call{SumColumns}{$\bs{P}_{\yhf}^*$}
    \item[]
    \item[// ---------- Calculate~$\Var{\ftest}{\pdf{\yhf|\ftest}}$, Equation~\eqref{eqn: cond_variance} ---------- //]
    \State~${\bs{K}}_{{\Ds}^{*}}$=\Call{PosteriorCovariance}{$\mathcal{GP}_{\fmod|\Ds},\Opstest$}
    \State~$j,h=1$
    \State~$\bs{Y}_{\text{HF},\mathbb{V}}=[\bs{y}_{\text{HF,support}},\bs{y}_{\text{HF,support}}]$
    \For{$\mu_1,v_1$ in~$\bs{m}^*,\bs{v}^*$}
        \State~$i=1$
        \For{$\mu_2,v_2$ in~$\bs{m}^*,\bs{v}^*$}
            \State~$k = {\bs{K}}_{{\Ds}^{*}}(i,j)$
            \State~$\nvar=$\Call{GetNoiseGP}{$\mathcal{GP}_{\fmod|\Ds}$}
            \State~$\bs{\Sigma}=[[v_1+\nvar,k]^T,[k,v_2+\nvar]^T]$ 
            \State~$\bs{\mu}=[\mu_1,\mu_2]^T$
            \State~$\bs{p}_{\yhf,\mathbb{V}^*}+=$\Call{Normal}{$\bs{Y}_{\text{HF},\mathbb{V}},\bs{\mu},\bs{\Sigma}$}
            \State~$i+=1$
            \State~$h+=1$
        \EndFor
        \State~$j+=1$
    \EndFor
    \State~$\bs{p}_{\yhf,\mathbb{V}^*}=\bs{p}_{\yhf,\mathbb{V}^*}/h-(\bs{p}_{\yhf,\mathbb{E}^*})^2$
    \item[]
    \State \Return~$\bs{p}_{\yhf,\mathbb{E}^*}$,~$\bs{p}_{\yhf,\mathbb{V}^*}$
\end{algorithmic}
\end{algorithm}

\FloatBarrier
\section{Appendix: Considerations of Computational Complexity, Costs and Speed-up for Numerical Relaxation}
\label{appendix: comp_cost}
In case the LF model is created by numerical relaxation, we want to provide brief considerations about the expected speed-ups. Again, we emphasize that numerical relaxation is only one way to create computationally cheaper model versions. More speed-ups can be achieved by employing simplified physical representations or geometries and combinations.
We investigate factors that influence the computational cost for approximating the solution of a nonlinear system of partial differential equations based on Galerkin-based discretization methods for transient problems with an iterative solution of nonlinear systems of equations within each time step.
Computational costs can be decomposed in contributions by the present number of degrees of freedom~(DoFs), i.e., the number of unknowns arising from the numerical discretization, the necessary number of iterations until convergence, as well as the efficiency of the implementation depending on the computational complexity of a chosen numerical algorithm but also the optimization level of a specific code. Thus, we obtain~\cite{fehn2018efficiency}:
\begin{align}
    \text{cost} \propto \text{DoFs}\cdot\text{time steps}\cdot \text{iterations}\cdot \frac{1}{\text{efficiency of implementation}}
\end{align}
We describe the general case of high-order discontinuous Galerkin methods~\cite{Hesthaven07} as these methods contain the widely used finite-element method and finite-volume method as a special case. The spatial dimensionality of the investigated problem is abbreviated by~$d$. The polynomial degree for the Ansatz functions is denoted by~$k$ and the measure for the element size by~$h$. The spatial discretization results in~$N_{\text{ele}}$ elements with~$(k+1)^d$ degrees of freedom~(and similarly~$k^d$ for a continuous finite element space) per element, assuming hexahedral elements and scalar fields. The number of elements is inversely proportional to~$h$ to the power of the dimension of the problem:~$N_{\text{ele}}\propto \frac{1}{h^d}$. The number of degrees of freedom~(DoFs) can then be summarized in Equation~\eqref{eqn: dofs}:
\begin{align}
\label{eqn: dofs}
    \text{DoFs} \propto N_{\text{ele}}\cdot (k+1)^d\propto \left(\frac{k+1}{h}\right)^{d}
\end{align}
The cost associated with time stepping is inversely proportional to the time step size~(we assume a mean time step for the cost considerations). Besides accuracy demands, the maximal possible time step size is constrained by the stability limits of the deployed solver. As a general discussion of the stability theory for arbitrary solvers is not expedient, we confine the analysis to solvers for transient fluid dynamics applications where the time step is selected according to the CFL condition~$(*)$, resulting in~\cite{fehn2018efficiency}
\begin{align}
\label{eqn: timestep}
    \text{time steps} \propto \frac{1}{\Delta t}\stackrel{^{(*)}}{\propto} \underbrace{\frac{k^\gamma}{h}}_{\substack{\text{CFL}\\ \text{relationship}}} \text{with} \ \gamma \in \left[1,2\right]\ ,
\end{align}
and argue that the resulting time step size is small enough to ensure a time-accurate solution for many problems. Computational costs related to the iterative solution of systems of equations for the unknown degrees of freedom are dependent on solver tolerances~$\boldsymbol{\epsilon}_{\text{solver}}$, and in general, also on the element size~$h$ as well as on the polynomial degree~$k$ of the Ansatz functions. For \emph{robust solvers}~$(**)$, e.g., multigrid, one can assume that the spatial discretization~($h,k$) does not influence iteration counts. A dependency of computational costs on the solver tolerance according to~$-\log\boldsymbol{\epsilon}_{\text{solver}}$ provides a good general model for solvers with optimal complexity so that we write:
\begin{align}
\label{eqn: iterations}
    \text{iterations} =f(\boldsymbol{\epsilon}_{\text{solver}},h,k)\stackrel{^{(**)}}{\approxprop} -\log\boldsymbol{\epsilon}_{\text{solver}}
\end{align}
Under \emph{efficiency of implementation}~(in DoFs computed per second), we imply the speed at which certain elementary operations of a PDE solver can be performed on given computer hardware and a given implementation. This factor is summarized in~$g(h,k)$ regarding the serial performance of a code, as well as the effects of parallel scalability that we summarize in the coefficient of parallel efficiency~$\eta_{\text{parallel}}(h,k)$. Furthermore, we introduce the speed-up through floating-point precision~$\mathfrak{p}\in \{1,2\}$, where~$\mathfrak{p}=1$ for~(standard) double precision and~$\mathfrak{p}=2$ in case the solver uses single floating-point precision. The factor~$g(h,k)$ mainly depends on the implementation variant used for the solver. We here focus on implementation strategies that have \emph{optimal complexity}~$(***)$ w.r.t. the polynomial degree~$k$ and mesh size~$h$ so that we can assume the serial performance to be almost independent of these parameters for~$k\leq 10$, see~\cite{kronbichler2019fast}. We further assume \emph{optimal parallel scalability}~$(***)$:
\begin{align}
\label{eqn: implementation}
\begin{split}
    \text{efficiency of implementation} &\propto g(h,k)\cdot\eta_{\text{parallel}}(h,k)\cdot \mathfrak{p}\\
    &\stackrel{^{(***)}}{\propto} \mathfrak{p}\ \text{  with } \mathfrak{p}=
    \begin{cases}
    1,&\text{for double precision}\\
    2,&\text{for single precision}\\
    \end{cases}
    \end{split}
\end{align}
The speed-up~$f_{\text{HF/LF}}$ by a numerical relaxation can then be expressed as:
\begin{align}
\label{eqn: speedup}
    f_{\text{HF/LF}}(k,d,h,\mathfrak{p}) :=\frac{\text{costs HF}}{\text{costs LF}}=\frac{\left(\frac{k_0+1}{h_0}\right)^{d_0}\cdot \frac{k_0^\gamma}{h_0}\cdot\frac{1}{\mathfrak{p}_0}\cdot(-\log\boldsymbol{\epsilon}_{\text{solver},0})}{\left(\frac{k+1}{h}\right)^{d}\cdot \frac{k^\gamma}{h}\cdot\frac{1}{\mathfrak{p}}\cdot (-\log\boldsymbol{\epsilon}_{\text{solver}})}
\end{align}

\begin{figure}[htbp]
\centering
\includegraphics[scale=0.22]{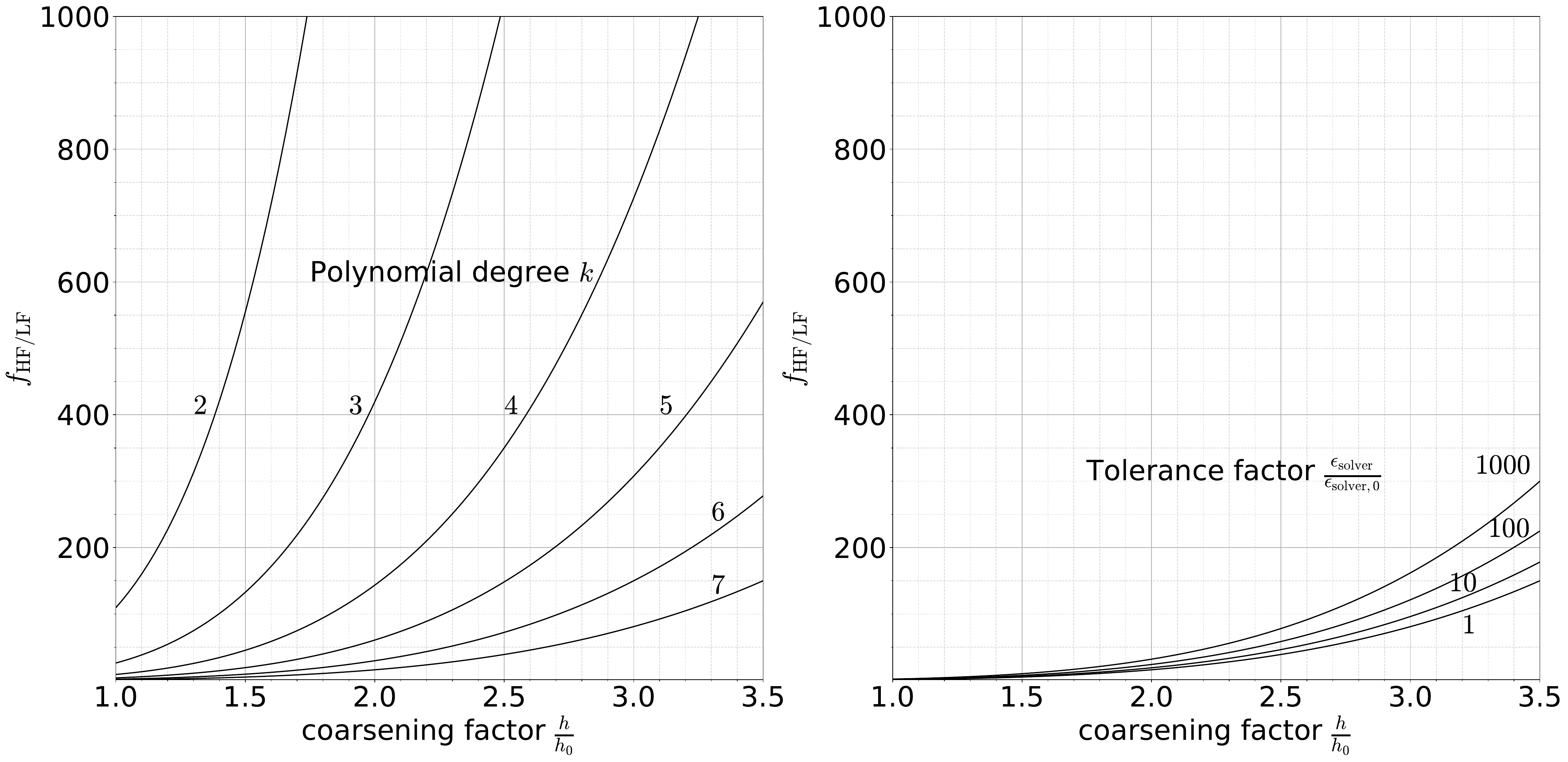} 
\caption{Example of theoretical speed-ups for a LF model, generated by numerical relaxation. Left: Speed-up over polynomial degree~$k$ and mesh-coarsening factor~$\frac{h}{h_0}$. The HF reference uses~$k_0=7,\ \epsilon_{\text{solver,0}}=10^{-6}$; Right: Speed-up over tolerance factor~$\frac{\epsilon_{\text{solver}}}{\epsilon_{\text{solver,0}}}$ and mesh-coarsening, with HF reference of~$k_0=2,\ \epsilon_{\text{solver,0}}=10^{-6}$. Along with the spatial coarsening a temporal relaxation is carried out according to the CFL constraint.}
 \label{fig: speed_up} 
\end{figure}

\Cref{fig: speed_up} illustrates potential speed-ups due to numerical relaxation of the original problem. According to the CFL condition, element or mesh coarsening is always shown in combination with a time step change. From a practical perspective, especially the change of the polynomial degree~$k$ for the Ansatz function of the Galerkin approximation can lead to large speed-ups even for the same mesh. In contrast, the relaxation of couplings or tolerances has a smaller impact~(logarithmic expression). An additional speed-up factor of two can be achieved when we relax the floating-point precision to single-precision. Even though the presented speed-ups are theoretical values, one can expect tremendous efficiency gains without the need for a completely new computational model.

\section{Appendix: Details Stochastic Flow Past a Cylinder}
\label{sec: details_cylinder}

Expression~\eqref{eqn: Dirichlet_BC} and~\eqref{eqn: custom_kernel} can be rewritten for easier implementation with available software packages such as \emph{GPy}~\cite{gpy2014}:
\begin{align}
\label{eqn: kernel_rewrite}
\begin{split}
\tilde{u}_x(y,t)&\sim\mu_u(y)\cdot \sin(\pi t/T)+0.125\cdot\mu_u(y)\cdot\mathcal{GP}\left(0,\kfun{u}{y}{y}\right)\cdot\sin(\pi t/T)\\
&=\mu_u(y)\left(1+0.125\cdot \mathcal{GP}\left(0,\kfun{u}{y}{y}\right)\right)\cdot \sin(\pi t/T)
\end{split}
\end{align}

Discrete realizations of the random inflow field can be computed using standard pseudo-random number generators. We define a vector of points~$Y_{\Gamma}$ on~$\Gamma_{D,u}^{\mathcal{F}}$ on which we evaluate the random inflow BC to yield the velocity vector~$U_{\Gamma}$. The Dirichlet boundary condition can then be imposed by a Galerkin projection step.
\begin{align}
\label{eqn: realizations_inflow}
\begin{split}
\mathbf{u}_{\Gamma}&=(\mathbf{m}+\mathbf{g})\cdot\sin(\pi t/T), \text{ with }\mathbf{m}=\mu_u(Y_{\Gamma})\text{ and}\\
\mathbf{g}&=\mathbf{L}\cdot\mathbf{r}\sim\mathcal{GP}\left(\boldsymbol{0},\kfun{u}{y}{y}\right),\ \text{ with  } \mathbf{r}\sim\nd{\mathbf{r}}{\boldsymbol{0},I} \text{ and  } \mathbf{L}\cdot\mathbf{L}^T=\Kmattest
\end{split}
\end{align}
In Equation~\eqref{eqn: realizations_inflow} the matrix~$\mathbf{L}$ denotes the Cholesky factorization of the covariance matrix~$\Kmattest=\kfun{u}{Y_{\Gamma}}{Y_{\Gamma}}$. The normally distributed vector~$\mathbf{r}$ has the dimension of~$Y_{\Gamma}$, which in our case was discretized by 200 points. The resulting stochastic dimension~(before dimensionality reduction) was~$d=203$. For subsequent parts of our analysis~(as put forth in \Cref{sec: aspects_implementation}), we can directly calculate a low dimensional representation~$\hat{\Xtest}$ of the high-dimensional inputs~$\Xtest$ by computing a truncated Karhunen-Lo\`eve expansion~(KLE)~(unsupervised dimensionality reduction) of the random field. The random variables~$\tilde{R},\ \tilde{\nu}$ and~$\tilde{y}_c$ are independent. Hence, their dimension cannot be further reduced. \Cref{fig: inflow_kle} shows on the right side realizations of the random inflow profile for~$t=T/2$~(solid lines) along with their truncated KLE approximation of order six~(dashed-lines). The bar chart~(left) shows the explained variance over the KLE truncation order. We decided to truncate the extension at order ten and store the reduced input data of the truncated inflow field and the three random variables as~$\hat{\Xtest}\in\real{\nsample\times13}$, in contrast to the original input data set~$\Xtest\in\real{\nsample\times203}$: 
\begin{figure}[htbp]
 \centering
  \includegraphics[scale=0.3]{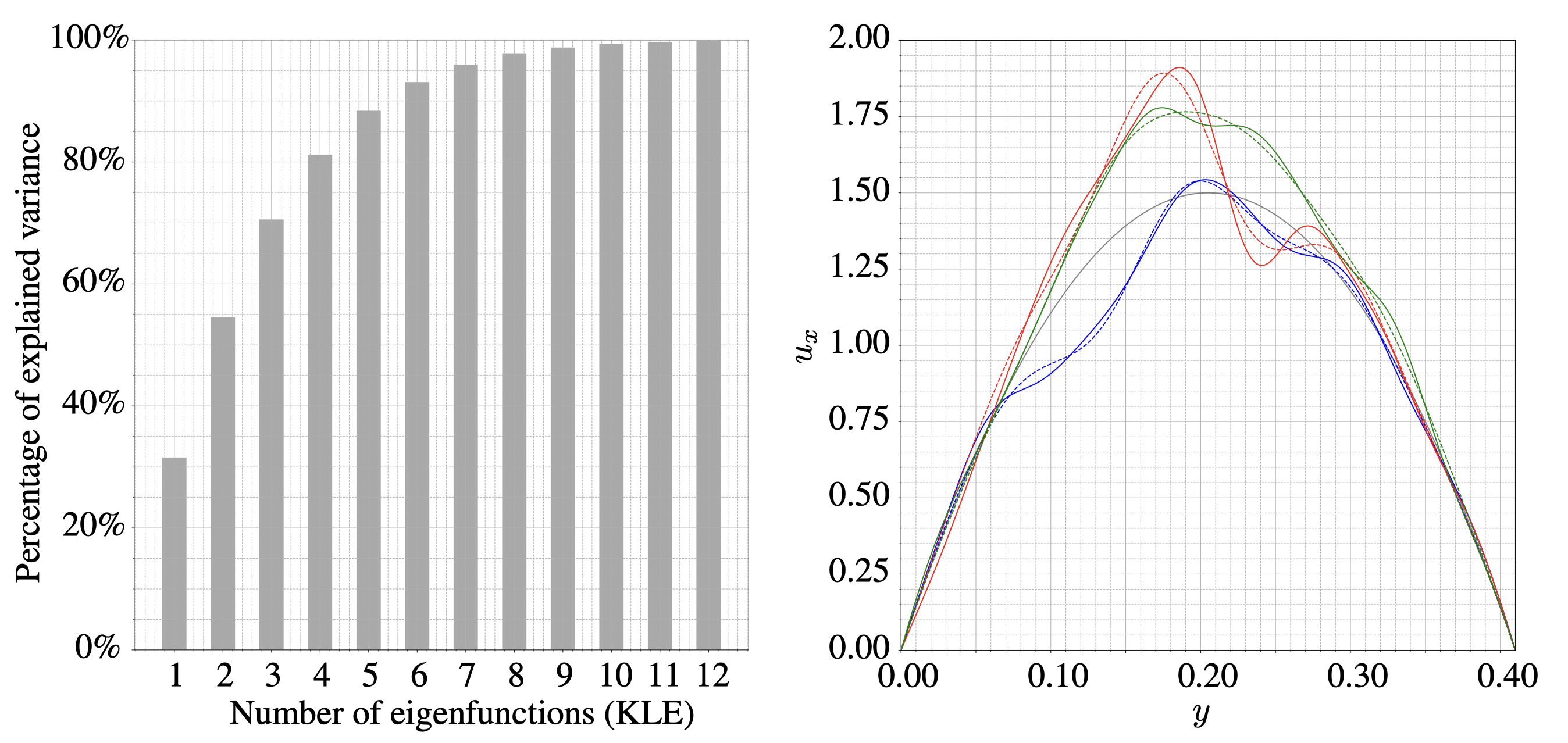}
  \caption{Left: Cumulative percentage of explained variance for a Karhunen-Lo\`eve expansion of the random inflow field; Right: Example samples~(solid line) of the random inflow field~$\tilde{u}_x(y)$ for~$t=T/2$ along with their Karhunen-Lo\`eve approximations~(dashed) of order six. The mean function of the random inflow is printed in light gray.} 
  \label{fig: inflow_kle}
 \end{figure}

 A summary of important properties for the stochastic flow problem is given in \Cref{tab: simulation_properties_2}.
\begin{table}[htbp]
  \caption{Properties used in the simulations~(random properties are printed bold)}
  \label{tab: simulation_properties_2}
  \centering
  \def\arraystretch{1.5}
  \begin{tabular}{lll}
    \toprule
    Property &Variable& Value\\
    \midrule
    Channel height&$H$&0.41\\
    Channel length&$L$&2.2\\
    \textbf{Lateral cylinder position}&$\tilde{y}_c$&$\ud{y}{0.16,0.24}$\\
    \textbf{Cylinder radius}&$\tilde{R}$&$\ud{R}{0.035,0.07}$\\
    \midrule
    \textbf{Kinematic viscosity}&$\tilde{\nu}$&$\ud{\nu}{9.5\cdot10^{-4},1.5\cdot 10^{-3}}$\\
    \textbf{Inflow BC}&$\tilde{u}_{x}(y,t)$&$\mu_u(y)\left[1+\frac{1}{8}\mathcal{GP}\left(0,\kfun{u_x}{y}{y}\right)\right]\sin(\frac{\pi t}{T})$\\
    Mean function at~$t=T/2$&$\mu_u(y)$&$U_{\rm{m}} \frac{4 y (H-y)}{H^2}$\\
    Mean max. velocity&$U_{\text{m}}$&1.5\\
    Correlation length scale&$\ls$&$0.08\cdot H$\\
    \bottomrule
  \end{tabular}
\end{table}
\FloatBarrier
\begin{remark}[Specifics of the employed discontinuous Galerkin code]
The code uses the matrix-free implementation developed in~\cite{kronbichler2019fast} which is available in the deal.II finite element library~\cite{dealII90}. On quadrilateral/hexahedral elements, the solution is approximated by tensor-product Lagrange polynomials of degree~$k \geq 2$ for the velocity unknowns, and degree~$k_p=k-1$ for the pressure unknowns for reasons of inf--sup stability. For efficient time integration, the method used in the present work relies on well-known projection methods that segregate the solution of velocity and pressure unknowns. We use a second-order accurate dual splitting scheme with an explicit treatment of the convective term to obtain a flow solver that is computationally efficient at high polynomial degrees~$k$. In all simulations, the same parameterized mesh according to~\cite{fehn2017stability} was used. 
\end{remark}

\FloatBarrier
\newpage
\section{Appendix: Details Stochastic Bending Wall in a Channel Flow}
\label{sec: details_bending_wall}

\begin{table}[htbp]
  \caption{Material and fluid properties used in the simulations~(uncertain properties are printed bold face)}
  \label{tab: simulation_properties}
  \centering
  \def\arraystretch{1.5}
  \begin{tabular}{lll}
    \toprule
    Property &Variable& Value\\
    \midrule
    Poisson ratio&$\nu^{\mathcal{S}}$&0.0\\    
    Solid density&$\rho^{\mathcal{S}}$&1.0\\
    \textbf{Young's modulus}&$\tilde{E}^{\mathcal{S}}$&$\lnd{E^{\mathcal{S}}}{\mu_E,\sigma_E^2}$\\
 &$\mu_E$&$6.392$\\
 &$\sigma_E^2$&$0.00498$\\
    \midrule
    Dynamic viscosity&$\mu^{\mathcal{F}}$&0.01\\
    Fluid density&$\rho^{\mathcal{F}}$&1.0\\
    \textbf{Inflow Field}&$\tilde{u}_{x}(y,z)$&$\mu_u(\bx)\left[1+\frac{1}{8} \mathcal{GP}\left(0,\kfun{u_x}{y}{y}\right)\right]$\\
    Mean function&$\mu_u(y,z)$&$0.05\cdot\left(1-\frac{4}{\left(h^{\mathcal{F}}\right)^2}y^2\right)\left(1-\frac{4}{\left(b^\mathcal{F}\right)^2}z^2\right)$\\
    Stationary covariance function&$\kfun{u_x}{y}{y}$&$\exp\left[\frac{-\left(y-{y}'\right)^2}{2 \ls^2}\right]$\\
    Correlation length scale&$\ls$&$0.08h^{\mathcal{F}}$\\
    \bottomrule
  \end{tabular}
\end{table}
\end{document}